\documentclass{pasj00}
\draft
\usepackage{lscape} %

\begin{document}
\SetRunningHead{J. Yokogawa et al.}{Discrete X-Ray Sources in the SMC}
\Received{2002/06/24}
\Accepted{2002/12/04}

\title{Review of Discrete X-Ray Sources in the Small Magellanic Cloud: 
Summary of the {ASCA} Results and Implication on 
the Recent Star-Forming Activity}
\author{%
Jun \textsc{Yokogawa},
Kensuke \textsc{Imanishi},
Masahiro \textsc{Tsujimoto}, 
Katsuji \textsc{Koyama}}
\affil{Department of Physics, Graduate School of Science,
Kyoto University, Sakyo-ku, Kyoto 606-8502}
\email{jun@cr.scphys.kyoto-u.ac.jp}
\email{kensuke@cr.scphys.kyoto-u.ac.jp}
\email{tsujimot@cr.scphys.kyoto-u.ac.jp}
\email{koyama@cr.scphys.kyoto-u.ac.jp}

\and

\author{%
Mamiko \textsc{Nishiuchi}}
\affil{Japan Atomic Energy Research Institute, 
Kansai Research Establishment, \\
8-1 Umebi-dai, Kizu-cho, Soraku-gun, Kyoto 619-0215}
\email{nishiuchi@apr.jaeri.go.jp}

\maketitle

\begin{abstract}
We made 22 observations on the Small Magellanic Cloud (SMC)
and covered full regions of the main body and the eastern wing
by the end of the {ASCA}  mission. 
We detected 106 discrete sources with a criterion of ${\rm S/N} >5$ and 
performed systematic analyses on all of the sources. 
We determined the source positions with 
an $\sim 40''$ error radius (90\% confidence) 
for sources detected 
in the central $20'$ radius of the GIS. 
We detected coherent pulsations from 17 sources. 
Among them, eight were newly discovered during this study. 
We classified most of these pulsars as X-ray binary pulsars (XBPs) 
based on their properties, such as the flux variability and 
the existence of an optical counterpart. 
We detected X-ray emission from eight supernova remnants (SNRs). 
Among them, five SNRs showed emission lines in their spectra, 
hence we regarded the five as thermal SNRs. 
We found that XBPs and thermal SNRs in the SMC
can be clearly separated by their spectral hardness ratio. 
Applying this empirical law to faint (thus unclassified) sources, 
we found 19 XBP candidates and four thermal SNR candidates. 
We also found several tens of candidates for 
active galactic nuclei,
both from the hardness ratio and the $\log{N}$--$\log{S}$ relation 
of extragalactic sources. 
Based on these {ASCA} results and further information from {ROSAT}, {SAX}, 
{RXTE}, {CGRO}, {Chandra}, and {XMM-Newton}, 
we compiled comprehensive catalogues of discrete X-ray sources 
in the Small Magellanic Cloud.  Using the catalogues, we derived the spatial 
distributions of XBPs and SNRs. 
XBPs and SNRs were found to be concentrated in the main body 
and eastern wing, which resembles the distribution 
of young stars with ages of $\sim 2 \times 10^7$~yr. 
By comparing  the source populations in the SMC and  our Galaxy, 
we suggest that the star-forming rate (per unit mass) in the SMC
was much higher than the Galaxy $\sim 10^7$~yr ago. 
We also discuss the recent change of the star-forming rate in the SMC. 
\end{abstract}

\section{Introduction}
Depending on the mass ($M$) and the existence of a companion, 
stars evolve with different sequences, come to deaths, 
and leave various remnants. 
Binary systems of low-mass stars ($M \lesssim 8M_\odot$) 
evolve through an accreting white dwarf and 
type-Ia supernovae (SNe), and make 
the remnants (SNRs) within $\sim 10^9$~yr 
\citep{Yoshii1996}. 
Massive stars ($8 M_\odot \lesssim M \lesssim 40M_\odot$), 
either single or binary, evolve more rapidly 
and come to type-II SNe, and leave SNRs within some $10^7$~yr 
(in this paper we designate all non--type-Ia SNe as type-II), 
leaving neutron stars (NSs). 
Crab-like pulsars or high-mass X-ray binaries (HMXBs) 
are the remnants of single or binary massive stars, respectively. 
More massive stars leave black holes (BHs) within some $10^6$~yr. 
Although the formation mechanisms of low-mass X-ray binaries (LMXBs) 
are still in dispute,  NSs in LMXBs and their companion stars  belong to
old populations with ages of $\gtrsim 10^{8-9}$~yr. 
The remnants of stars, i.e., 
supernova remnants from type-Ia and type-II SNe 
(hereafter, type-Ia SNRs and type-II SNRs, respectively), 
young SNRs, Crab-like pulsars, HMXBs, LMXBs, and BH binaries comprise
the brightest  X-ray sources in a galaxy 
with a luminosity of $\gtrsim 10^{35}$~erg~s$^{-1}$. 
%
Therefore, bright X-ray sources in a galaxy 
carry much information on the past star-forming activity, 
such as the rate and the site of star formation, mass function, 
and binary frequency.

Each  class of the stellar remnants distinguishes itself  
by the X-ray  spectrum and the temporal behaviour. 
Since X-ray emission from SNRs is mainly attributable to a 
shock-heated plasma, their spectra are relatively soft ($\lesssim 2$~keV) 
and dominated by emission lines from highly ionized atoms. 
The compositions of the atomic species produced  by SNe 
depend on the types: 
light elements such as O, Ne, and Mg are mainly ejected from 
type-II SNe, while heavier elements such as Si, S, Ar, Ca, and Fe 
come  from type-Ia SNe. 
Sources in the other classes have harder spectra than SNRs,
showing a significant flux even at energies $\gtrsim 10$~keV. 
LMXBs have characteristic spectra which can be represented by 
a two-component thermal model
\citep{Mitsuda1984}.  
Some of them  occasionally  exhibit  X-ray bursts.
BH binaries experience spectral transitions 
between the high-soft and low-hard states. 
Crab-like pulsars exhibit 
coherent pulsations with a short period ($\lesssim 1$~s) and 
a monotonous increase of the spin period (except for glitches). 
HMXBs have the hardest spectra, and exhibit 
long-term flux variations with a factor of $\gtrsim 10$. 
Many HMXBs  exhibit coherent pulsations with 
a long period ($\gtrsim 1$~s), and are thus classified as 
X-ray binary pulsars (XBPs). 
The {ASCA} satellite \citep{Tanaka1994}, with a reasonable
energy resolution and high sensitivity for hard X-rays ($> 2$~keV) 
and a fine timing resolution, had a capability to distinguish
the above-mentioned variety of X-ray properties, 
and to classify the X-ray source populations.

The Small Magellanic Cloud (SMC), a satellite of our Galaxy, is the
next-nearest neighbor after the Large Magellanic Cloud (LMC).  The
proximity (60~kpc is assumed in this paper; 
van den Bergh 2000), 
reasonable angular size ($\sim 3^\circ \times 3^\circ$), and low
interstellar absorption in the direction of the SMC are all favorable
for an unbiased survey of the X-ray source populations covering  the
entire galaxy.  Surveys of soft X-ray sources (below $\sim 2$~keV)
have been carried out with the {Einstein} and {ROSAT} satellites 
\citep{Wang1992,Haberl2000a,Sasaki2000}. 
\citet{Haberl2000a} 
present the most complete catalogue, which contains 517 {ROSAT} PSPC 
(Position Sensitive Proportional Counter) sources. 
\citet{Sasaki2000} 
used {ROSAT} HRI (High Resolution Imager) 
to determine the most accurate positions of 121 sources. 

A hard X-ray study with {ASCA} 
in our earlier survey \citep{Yokogawa2000e}
classified many new sources into XBPs and thermal SNRs, 
and provided a simple (but reliable) method for this 
classification. 
However, because the {ASCA} survey did not
cover all of the SMC fields, it may have had some bias for 
the population and distribution study.

After this early study, we have  carried out new {ASCA} observations,
and we have now covered the entire region of the SMC. 
This paper reports on the summary with  particular emphasis 
on the {ASCA} new results.  However, for completeness, we 
also extend  to all the {ASCA}  results as well as
to some related results from {ROSAT}, {SAX}, 
{RXTE}, {CGRO}, {Chandra}, and {XMM-Newton}. 
Observation fields and the method of data reduction 
are presented in section 2. 
Source detection and position determination are 
described in section 3; the positional accuracy is discussed in detail. 
Timing and spectral analyses are performed for all sources 
as described in section 4. 
Comments on X-ray pulsars and SNRs are presented in section 5. 
Pulsar statistics, source classifications, 
source populations, and spatial distributions are 
discussed in section 6.

\section{Observations and Data Reduction}
\begin{figure}
 \FigureFile(80mm,80mm){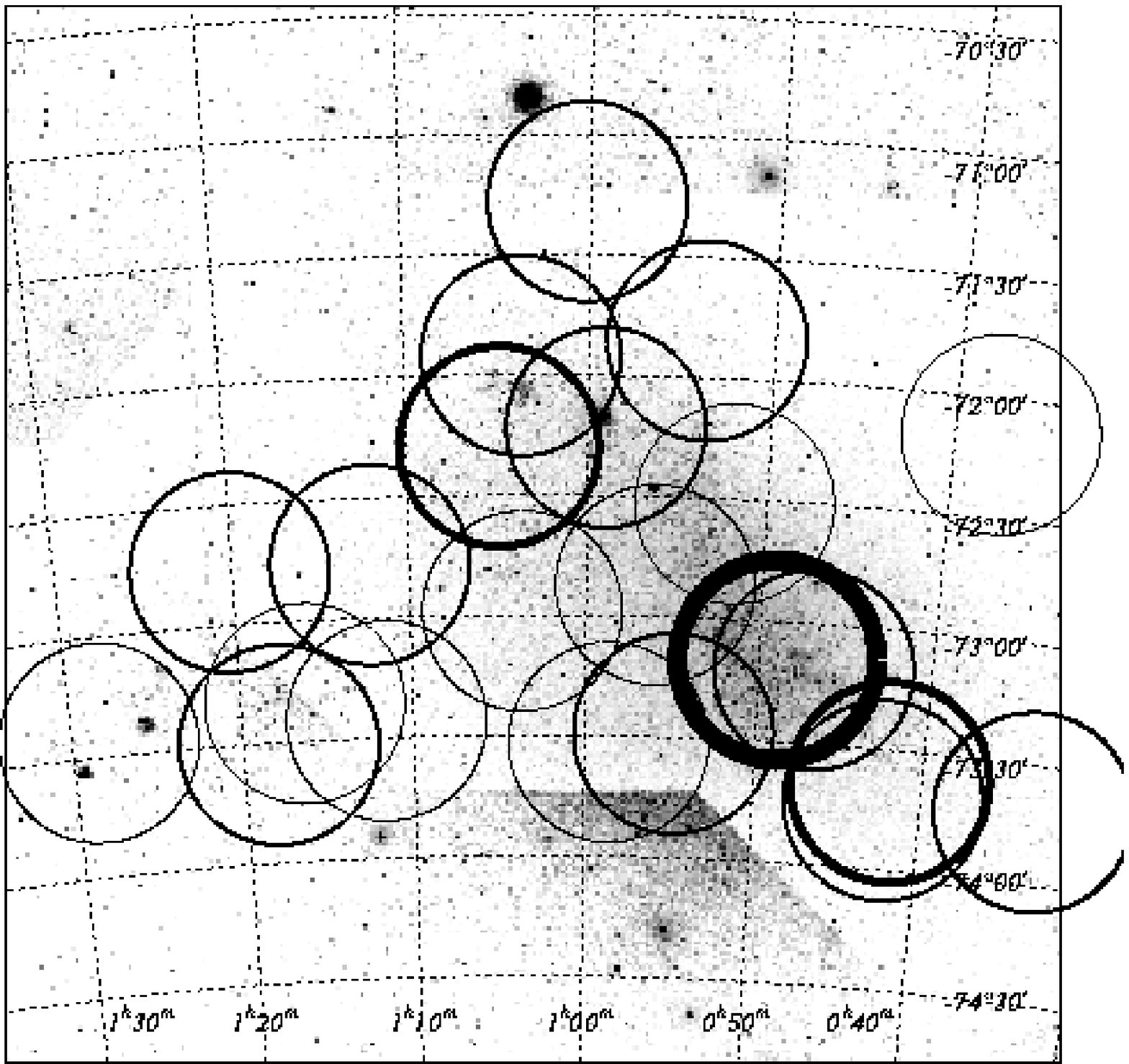}
 \caption{All observation fields with {ASCA} GIS superimposed on the optical image 
of the SMC region from the Digitized Sky Survey (DSS). 
Equatorial coordinates with an equinox of 2000 are also shown. 
Each observation field is represented by a $50'$-diameter circle, 
the thickness of which is proportional to the exposure time. 
The discontinuity seen in the DSS image is an artifact. 
  \label{fig:obsfield}}
\end{figure}

{ASCA} observed the 22 SMC regions by the end of the mission, 
as summarized in table \ref{tab:obs}. 
Although the observations made before 1999 aimed at specific objects 
such as SNRs, X-ray pulsars, and a supersoft X-ray source, 
the assembly of these observations had already covered 
most of the main body and the eastern wing of the SMC. 
Our earlier study \citep{Yokogawa2000e} is based on 
these observation data 
(except for obs.\ J). 
In order to cover all of the blank area, 
a survey project (SMC~1--10 in table \ref{tab:obs}) 
including long-exposure observations 
(SMC~SW~N1 and N2 in table \ref{tab:obs}) was performed. 
These observations covered most of the SMC region, 
as shown in figure \ref{fig:obsfield}.
In this study, we used all of these observation data 
and carried out various analyses in a coherent manner.

In each observation, X-ray photons were collected with 
four XRTs (X-Ray Telescopes; Serlemitsos et al.\ 1995)
and detected separately with two GISs 
(Gas Imaging Spectrometers; Ohashi et al.\ 1996)
and two SISs 
(Solidstate Imaging Spectrometers; Burke et al.\ 1994). 
We rejected any data obtained in the South Atlantic Anomaly, 
or when the cut-off rigidity was lower than 4~GV, 
or when the elevation angle was lower than 5$^\circ$. 
Particle events in the GIS data 
were removed by the rise-time discrimination method.
SIS data obtained when the elevation angle from the bright Earth was lower 
than 25$^\circ$, 
or with hot and/or flickering pixels, 
were also rejected. 
The effects of RDD 
(Residual Dark Distribution) 
on the SIS data were corrected 
with the method given in \citet{Dotani1997}
for observations carried out later than 1996. 
After the screening, 
the total available exposure time for two GISs was $\sim 2000$~ks.

In order to uniformly study X-ray source populations, 
the GIS is more suitable than the SIS 
because of its larger field of view, 
larger effective area at high energy, 
and better time resolution. 
Therefore, we mainly used the GIS data in this study, 
while the SIS data were used for peculiar objects 
which need better energy resolution 
and/or better spatial resolution.

\section{Source Catalogue}

\subsection{X-ray Images}
Images in each observation were constructed 
in the sky and detector coordinate systems 
(hereafter ``sky images'' and ``detector images'', respectively), 
with the {\sf XSELECT} package. 
In the sky image, the position of each photon is determined 
using the instantaneous satellite attitude at the {\it incident time} of each 
photon.  Therefore, the sky image is properly corrected for the image blurring 
due to attitude flickering. The data processing for the sky image is, 
however, limited to the photons coming within $\sim 20'$ radius on the GIS 
center.
In the detector image, photons are accumulated in the coordinates fixed to each 
detector, and are then converted to the sky coordinates using 
the {\it average} attitude of the satellite during the observation. 
This software technique can be applied to a larger FOV  of $\sim 25'$ in radius.
We  properly used the two different images: the sky images  for sources within 
the central $\sim 20'$ radius (hereafter ``inner circle''), and the detector 
images for  the concentric region of $\sim 20'$--$25'$ 
(hereafter ``outer ring''). 
The outer ring region has a higher background,  larger calibration error and  
distortion of the PSF than the inner circle region.
For SIS, we always used sky images.

\begin{figure}
 \FigureFile(100mm,100mm){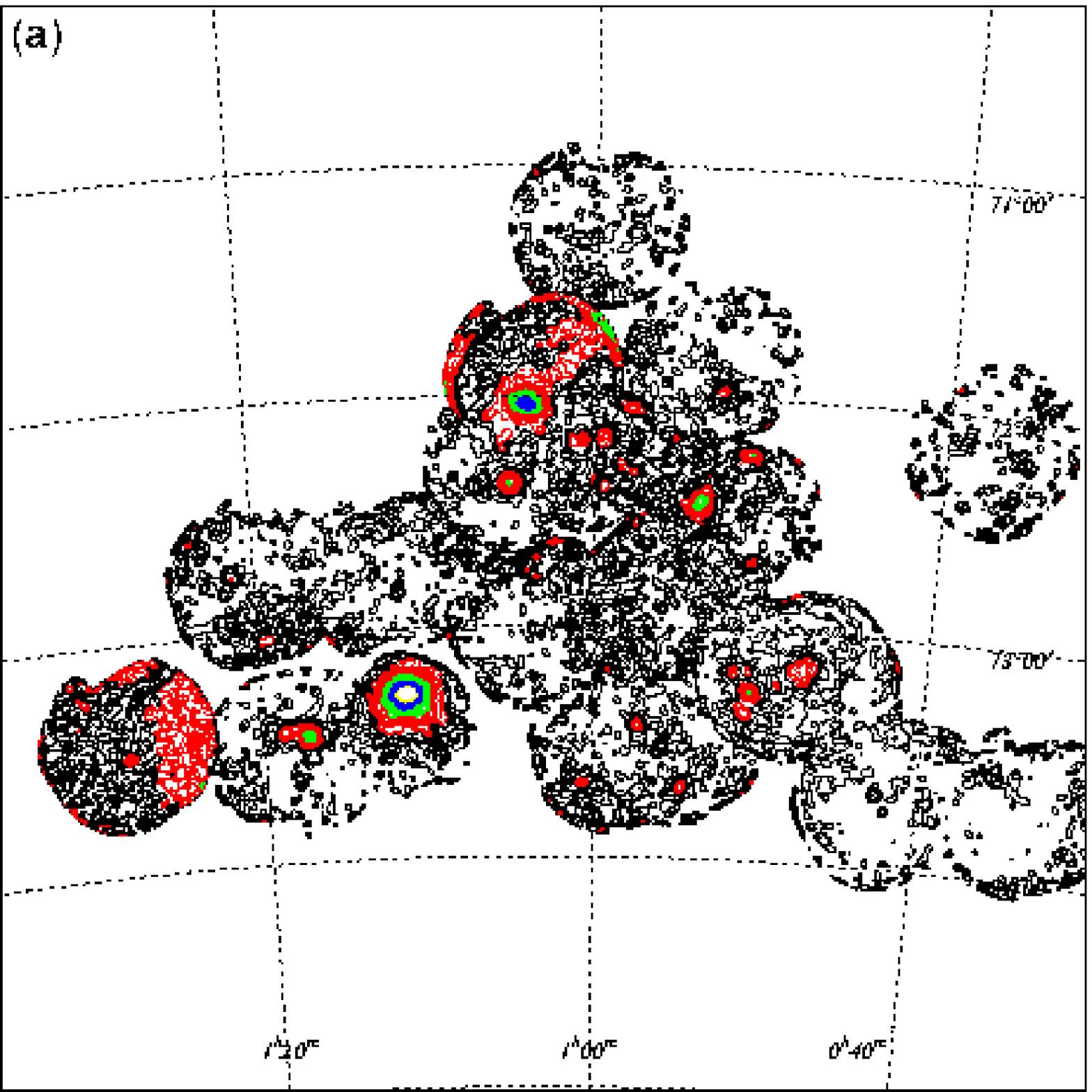}

 \FigureFile(100mm,100mm){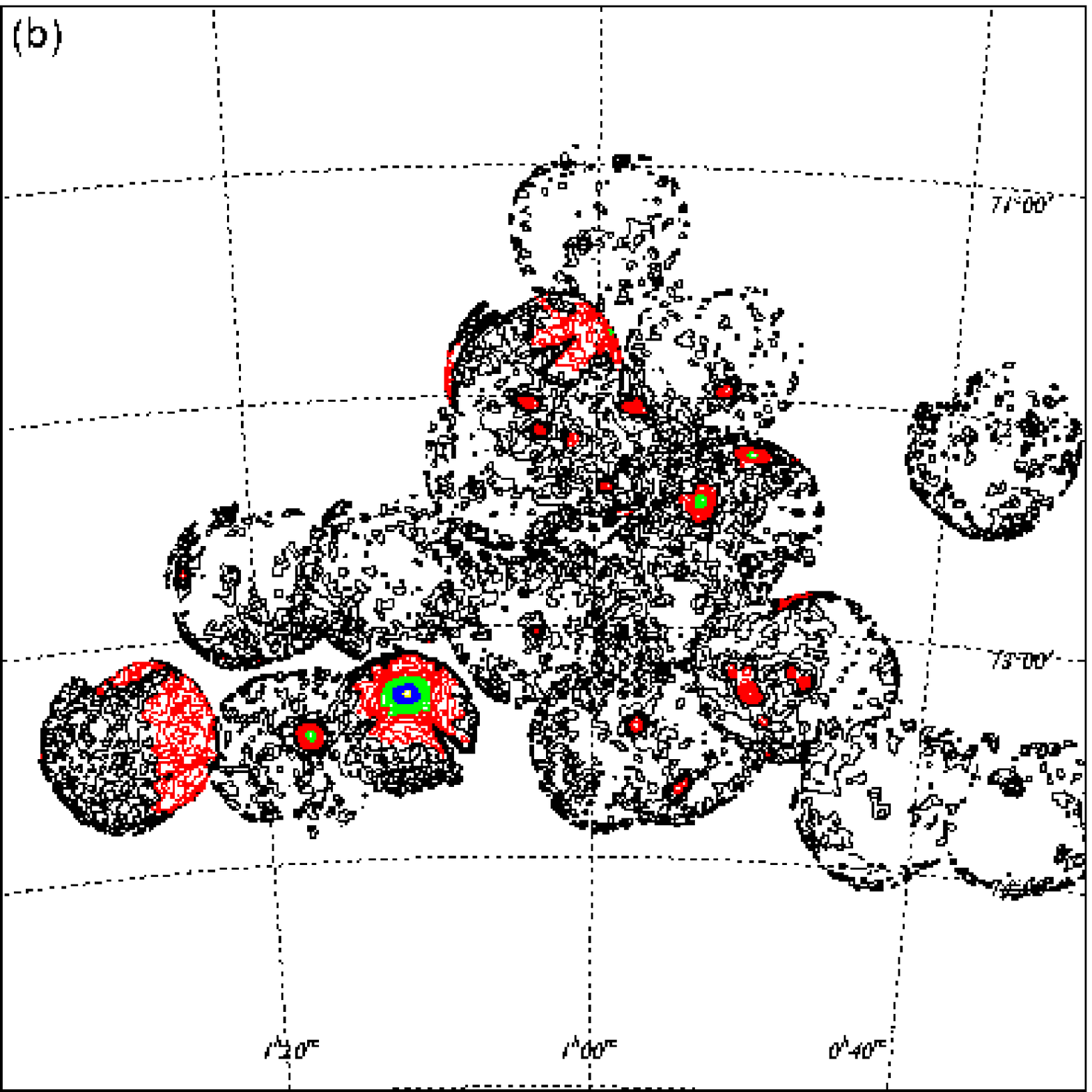}
 \caption{Mosaic images of the SMC obtained with {ASCA} GIS 
in the soft (a: 0.7--2.0~keV) and hard (b: 2.0--7.0~keV) bands, 
overlaid with equatorial coordinates (J2000). 
The effects of non--X-ray background, telescope vignetting, 
and difference of exposure time between observations were corrected. 
Contour levels are linearly spaced. 
Of the two observations centered on SMC~X-1 (observations A and C), 
only observation C was used, in which SMC~X-1 was much fainter.
\label{fig:mosaic}}
\end{figure}

\begin{figure}
 \FigureFile(80mm,80mm){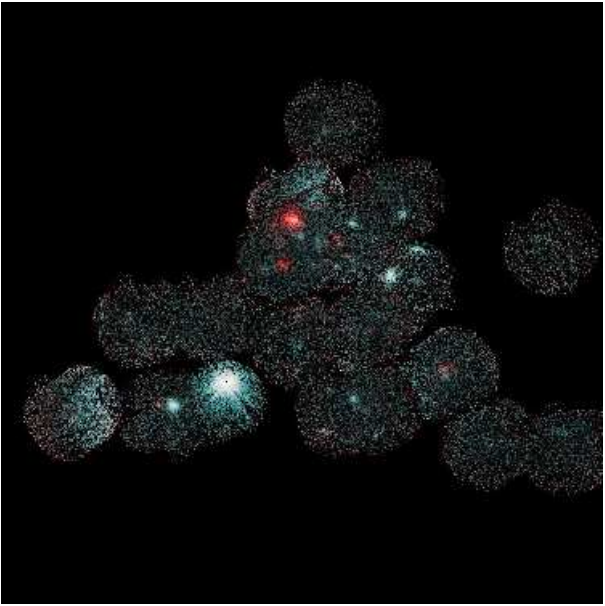}
 \caption{
Two-color X-ray mosaic image of the SMC. 
Red and blue indicate X-ray photons in the 
soft (0.7--2.0~keV) and 
hard (2.0--7.0~keV) bands, respectively. 
\label{fig:mosaic_color}}
\end{figure}

Figure \ref{fig:mosaic} shows 
the GIS mosaic images in the soft (0.7--2.0~keV) and 
hard (2.0--7.0~keV) bands, 
created according to the method developed by \citet{Sakano2000}. 
A color-coded image in which soft and hard photons are 
indicated by red and blue is given in figure \ref{fig:mosaic_color}. 
In the color image, 
many hard sources and a few soft sources are clearly found.

\subsection{Source Detection\label{sec:extract}}
For each observation, X-ray sources were extracted from images 
in the soft (0.7--2.0~keV), hard (2.0--7.0~keV), and 
total (0.7--7.0~keV) bands. 
We smoothed the images with a Gaussian filter ($\sigma = 30''$), 
and examined the significance of each local peak (source candidate) 
found in the images as follows. 
Photons were extracted from a circle of $3'$ radius 
centered on the peak, 
in which 90\% of the incident photons were contained 
\citep{Serlemitsos1995}, 
or from an ellipse at larger off-axis angles 
because of the distortion of the PSF. 
In several cases, a smaller circle/ellipse 
was used to avoid contamination from nearby peaks. 
For observation\ J, a larger circle was used, 
because in this observation 
the spatial resolution was reduced by a factor of 4, 
and thus the images were blurred. 
These photon events were also used in subsequent analyses 
described in the following subsections. 
%
Background regions were selected from the sky near each peak. 
%
We then derived the ${\rm S/N}$, defined as 
%
%
${\rm S/N} \equiv [n({\rm P}) - n({\rm B})]/\sqrt{n({\rm B})}$,
where $n({\rm P})$ and $n({\rm B})$ represent the photon counts 
in the circle/ellipse at the peak and 
in the background region, respectively. 
The local peak was identified as an X-ray source 
if the ${\rm S/N}$ ratio exceeded $5$ 
in at least one of the soft-, hard-, or total-band images. 
In all, we detected 106 sources, 
of which 21 were detected in multiple observations  
(subsubsection \ref{sec:asca-asca}).

%

In observation\ Q, sources No.\ 26 and No.\ 27 were resolved 
only in the SIS image with a separation of $\sim 1\farcm 3$. 
No.\ 85 was detected only in the hard band 
because of severe contamination from No.\ 81 in the soft band. 
No.\ 88 and No.\ 89 are located near  
the calibration isotope of GIS~3; we thus used only GIS~2 
to estimate the significance. 
Since No.\ 88 and No.\ 89 are separated by only $\sim 1\farcm 5$, 
which caused severe mutual contamination, 
we used very small circles to estimate their ${\rm S/N}$ ratio. 
We found that the ${\rm S/N}$ of No.\ 89 well exceeds 5, 
while that of No.\ 88 is slightly less than 5. 
However, a local peak at No.\ 88 was 
also found evidently in GIS~3 
(although no quantitative estimation is possible); 
we thus regard No.\ 88 as an X-ray source.

\subsection{Position Determination\label{sec:position}}
\subsubsection{Absolute accuracy of the position\label{sec:absolute}}
We first determined the position of each source 
simply by the coordinates of the peak pixel in the smoothed GIS images, 
and then performed a correction developed by \citet{Gotthelf2000}. 
This correction compensates for the positional uncertainty 
caused by the {ASCA} attitude error, 
which depends on the temperature of the base-plate of the star-tracker 
and the geometry of the solar illumination. 
SIS images were used only for resolving sources No.\ 26 and No.\ 27 
in obs.\ Q. 
According to this correction, 
the coordinates of some X-ray pulsars with ``AX~J'' names 
(which have been included in previous publications) 
are now inconsistent with the source name; 
for example, the coordinates of No.\ 40 = AX~J0051.6$-$7311 
\citep{Yokogawa2000b} are now 
(00{$^{\rm h}$}51$^{\rm m}$44$\fs$5, $-73^\circ 10'34''$). 
In this paper, we do not rename these sources to avoid name confusion, 
and adopt the names used in the first publications for each pulsar.

This correction 
reduces the systematic positional uncertainty to $24''$ 
(90\% error radius) for sources detected in the 
central $10'$ radius of the GIS \citep{Gotthelf2000}. 
However, additional errors from 
the photon statistics and the method of position determination, 
and errors 
for the sources located out of the central $10'$ radius, 
are unknown. 
Therefore, we examined the ``practical'' errors 
for sources detected anywhere in the GIS as follows. 

So far, the {ROSAT} HRI catalogue \citep{Sasaki2000}
presents the most accurate positions for the SMC X-ray sources, 
with an error radius of $\sim 1''$--$10''$. 
Several sources in the {ROSAT} PSPC catalogue \citep{Haberl2000a}
also have a small error radius of $\lesssim 10''$. 
Therefore, we investigated the separation angles 
between the {ROSAT} sources and their {ASCA} counterparts, 
which would represent the ``practical'' errors for the {ASCA} sources. 

We primarily selected {ASCA} counterparts for the {ROSAT} sources 
that were within $90''$ of each {ROSAT} source. 
In order to reject accidental coincidences 
and ambiguous counterparts, 
we further employed spectral and temporal information of these sources 
as follows: 
(1) For sources catalogued in both the PSPC and HRI catalogues, 
      only HRI sources were used, which provide more accurate positions. 
(2) Only {ROSAT} sources with an error radius smaller than 
      $7''$ were used. 
      Since the error radii for {ASCA} sources are $> 24''$, 
      the additional error from the {ROSAT} sources 
      is $< 1''$ when we take a root-sum square of all the errors. 
(3) For a {ROSAT} source with a Be star or a supergiant companion, 
      the {ASCA} counterpart should exhibit coherent pulsations 
      with a period of $\gtrsim 1$~s. 
      The procedure for pulse detection is described in subsection \ref{sec:timing_ana}. 
      This criterion selects well-established XBPs. 
      Although {ASCA} source No.\ 94 in obs.\ C exhibited no significant 
      pulsations, 
      it entered an eclipse phase as the ephemeris predicts for SMC~X-1 
      \citep{Wojdowski1998}, 
      thus we regard No.\ 94 in obs.\ C as SMC~X-1. 
(4) For a {ROSAT} source at the position of a radio SNR, 
      the {ASCA} counterpart should exhibit a soft spectrum 
      with emission lines from ionized atoms. 
      The method used to detect emission lines is described in 
      subsubsection \ref{sec:spec_snr}. 
      This criterion selects bright thermal SNRs.

\begin{figure}
 \FigureFile(80mm,80mm){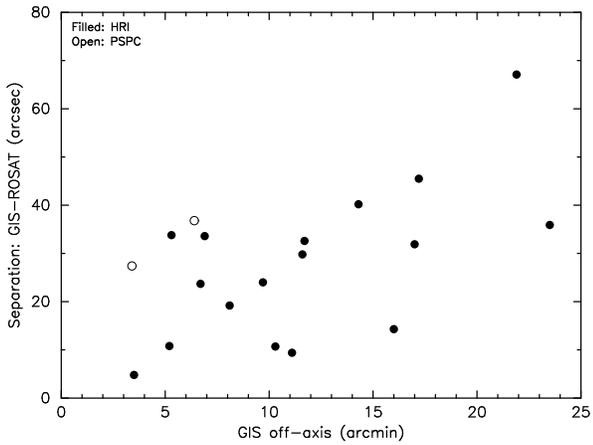}
 \caption{Plot of the separation angles between the 
{ROSAT}--{ASCA} counterparts as a function of the off-axis angle 
of the {ASCA} sources (see table \ref{tab:counter_abs}). 
The filled and open circles indicate the sources detected with 
{ROSAT} HRI and PSPC, respectively.\label{fig:sep-offax}}
\end{figure}

According to these criteria, we selected 
19 pairs of {ASCA}--{ROSAT} counterparts, 
as summarized in table \ref{tab:counter_abs}. 
Although No.\ 67 is certainly the counterpart for RX~J0059.2$-$7138 
(see subsubsection \ref{sec:2.7s}), 
this pair is not included in table \ref{tab:counter_abs} 
because No.\ 67 is detected at the very edge of the GIS 
(or may be slightly outside of the GIS) 
and so the position determination is not reliable. 
We show the separation angles as a function of the off-axis angle 
of the {ASCA} source in figure \ref{fig:sep-offax}. 
No clear correlation between the separation angle 
and the off-axis angle could be found. 
Out of 17 {ASCA} sources detected in the inner circle 
(off-${\rm axis} < 20'$), 
15 sources have separation angles less than $40''$. 
Therefore, we tentatively conclude that 
the ``practical'' error radius for GIS sources 
detected in the inner circle 
is $40''$ at 90\% confidence level, 
although the statistics are rather limited. 
This is similar to the result obtained from 
the more elaborate analysis by \citet{Ueda1999}. 
For sources detected in the outer ring, 
no constraint could be obtained due to the paucity of sources.

From the {ROSAT} and {Einstein} catalogues 
\citep{Haberl2000a,Sasaki2000,Wang1992}, 
we selected the counterpart for each {ASCA} source 
within a circle of a radius 
$\sim 60''$ for sources detected in the inner circle, 
or 
within a circle of a radius 
$\sim 70''$ for sources detected in the the outer ring. 
Radii larger than the 90\% error radius ($40''$) 
were used in order to simply avoid 
missing identification.

\subsubsection{Identification of sources detected in multiple {ASCA} observations\label{sec:asca-asca}}
As shown in figure \ref{fig:obsfield}, 
neighbouring {ASCA} observation fields more or less overlap each other. 
Therefore, a pair of detections found in two observations 
within the overlapped region may be 
from the same source.
In order to examine whether these pairs are the same source or not, 
we primarily selected pairs of detections within $90''$ of each other, 
and classified them into four classes (a)--(d) as follows: 
(a) Both of the sources exhibit coherent pulsations 
               with nearly the same period, or exhibit emission lines 
               from the same elements and have the same radio SNR 
               as a counterpart
               (see subsection \ref{sec:timing_ana} and subsubsection \ref{sec:spec_snr} for 
               the relevant analyses). 
(b) Both of the sources have soft spectra and 
               have the same radio SNR as a counterpart. 
               Pairs of No.\ 94 in obs.\ A and C and that in obs.\ I and C 
               (SMC~X-1) are also included in this class. 
               Classes (a) and (b) surely consist of pairs of XBPs 
               and thermal SNRs. 
(c) Both of the sources are located near the same pulsar 
               and their spectral parameters are consistent with 
               those of the pulsar. 
               Sources of class (c) are likely to be X-ray pulsars. 
(d) The remainder. 

\begin{figure}
 \FigureFile(80mm,80mm){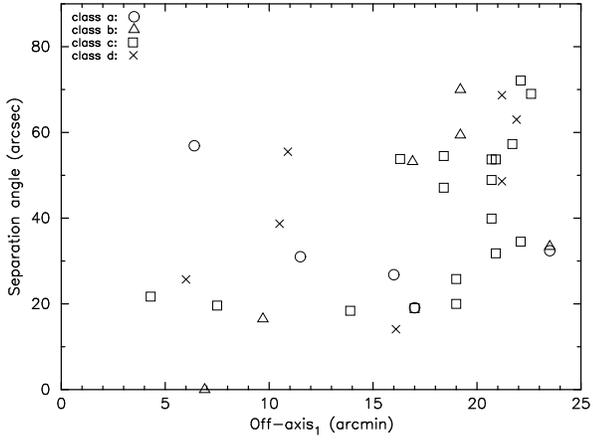}
 \caption{Plot of the separation angles 
as a function of the larger off-axis angle 
(off-axis$_1$ in table \ref{tab:asca-asca}) of 
{ASCA} sources detected multiple times. \label{fig:asca-asca}}
\end{figure}

We regarded detections in classes (a)--(c) as being from the same source, 
i.e., sources detected in multiple observations, 
and thus labeled them with the same source number 
in the {ASCA} catalogues (tables \ref{tab:cat_counter} and 
\ref{tab:cat_param}). 
We summarize the separation angle and the off-axis angles 
of classes (a)--(c) in table \ref{tab:asca-asca}, 
while in figure \ref{fig:asca-asca} 
we give a plot of the separation angle vs.\ the larger off-axis angle. 
We found that the separation angle is $\lesssim 60''$ 
if both of the two sources are detected in the inner circle, 
or $< 73''$ 
if at least one of the two is detected in the outer ring. 
Therefore, 
we regard 
pairs of detections in class (d) to be the same source 
if they satisfy the above condition, 
and labeled them the same source number.  
Pairs thus selected are also summarized in table \ref{tab:asca-asca} 
and plotted in figure \ref{fig:asca-asca}. 
After this selection, we concluded that 
{ASCA} detected 106 sources with no double count.

\section{Analyses on Discrete Sources}
In order to examine the nature of each source, 
we performed timing and spectral analyses in a coherent manner. 
The procedure of the analyses is essentially identical 
to that in \citet{Yokogawa2000e}. 

\subsection{Timing Analyses\label{sec:timing_ana}}
%
%
We performed a Fast Fourier Transform (FFT) analysis 
on all of the sources to search for coherent pulsations.
At first, for sources with high count rates, 
we used only high-bit rate data in order to utilize
the maximum time resolution (up to 62.5~ms). 
We then used high-bit and medium-bit data simultaneously 
for all sources, in order to achieve better statistics
at the sacrifice of the time resolution to 0.5~s 
(7.8125~ms for obs.\ J and 
125~ms for obs.\ O; see the caption of table \ref{tab:obs}). 
We detected coherent pulsations from 17 sources,
eight of which are new discoveries from this study. 
Examples of the power spectrum densities (PSDs) 
are already shown in figure~2 of \citet{Yokogawa2000e} 
or chapter~5 of \citet{Yokogawa2002b}. 
%
The detection of pulses from 
No.\ 26 (AX~J0049$-$732) and
No.\ 83 (AX~J0105$-$722) 
was not straightforward because of contamination from nearby sources. 
The details are described in 
subsubsection \ref{sec:9s} and subsubsection \ref{sec:3.34s}, respectively. 

In any observation, 
photon events were originally counted 
with a time spacing of 1/16 of the nominal resolution 
and stored in temporary memory. 
The events were then 
collectively sent to the telemetry 
with a time spacing equal to the nominal resolution. 
Therefore, if the event rate was so low as 
not to fill 
the memory, 
the time resolution could be 1/16 of the nominal value 
\citep{Hirayama1996}. 
For this reason, 
we carried out FFT analysis on several faint sources 
with a time resolution of 31.25~ms, 
using the high- and medium-bit data. 
The 87~ms pulsations from AX~J0043$-$737 were thus discovered 
(see subsubsection \ref{sec:87ms} for further details).

In order to determine the pulse period precisely, 
we performed an epoch folding search 
for the 17 sources from which pulsations were detected 
by FFT analysis. 
The orbital Doppler effect was corrected 
only for SMC~X-1, using the ephemeris presented by 
\citet{Wojdowski1998}. 
The derived pulse periods are presented in table \ref{tab:pulsars}.

We detected no pulsations by FFT analysis 
from three sources that are positionally coincident with 
known pulsars: 
No.\ 43 (RX~J0052.1$-$7319), 
No.\ 51 (XTE~J0055$-$724), 
No.\ 74 (RX~J0101.3$-$7211), and 
No.\ 94 in obs.\ C (SMC~X-1). 
Therefore, we performed an epoch folding search 
around the known periods. 
Since SMC~X-1 was in the 0.6-d eclipse phase 
during obs.\ C, we only used the data from the noneclipse times. 
Consequently, 
we detected a weak peak only from No.\ 51 at the known period of $\sim 59$~s, 
which is the evidence that No.\ 51 and XTE~J0055$-$724 
are the same source. 
This period is, however, not presented in table \ref{tab:pulsars} 
because of the low significance of the pulse detection.

We also searched for burst-like activities by using light curves 
binned with various time scales from $\sim 1$~s to $\sim 1$~hr. 
Although no source exhibited bursts typical of LMXBs, 
No.\ 20 (RX~J0047.3$-$7312 = IKT1, in obs.\ Q) showed a flare with a decay time 
of $\sim 2 \times 10^4$~s. 
Details are given in subsubsection \ref{sec:IKT1}.

\subsection{Spectral Analyses\label{sec:spec_ana}}
\subsubsection{Overview}
We analyzed the spectrum of each source 
and derived various parameters, as given in table \ref{tab:cat_param}: 
the hardness ratio (HR), photon index ($\Gamma$), 
temperature ($kT$), column density ({$N_{\rm H}$}), 
flux ({$F_{\rm X}$}), and absorption-corrected luminosity ({$L_{\rm X}$}). 

%
%
The analyses were not performed for 
No.\ 85, No.\ 88, and No.\ 89 because of severe contamination 
of these sources (see subsection \ref{sec:extract}). 
%
Spectra from GIS~2 and GIS~3 were co-added to increase the statistics, 
except for sources detected near the calibration isotope 
of either GIS and sources detected in only a single GIS\footnote{%
Since the FOVs of the two GISs are pointed toward slightly different 
directions, it is possible for 
a source to be located at the very edge of one GIS 
and outside of the other GIS. }. 
SIS spectra (SIS~0 $+$ SIS~1) were used 
for No.\ 26 and No.\ 27 in obs.\ Q in order to spatially resolve 
these sources, 
and for SNRs 
0045$-$734 (No.\ 21), 
0047$-$735 (No.\ 25), 
0057$-$7226 (No.\ 66), 
0102$-$723 (No.\ 81), and 
0103$-$726 (No.\ 82) 
to perform high resolution spectroscopy%
\footnote{For 0047$-$735, the SIS spectrum in obs.\ F was not used 
because the statistics were too poor.}. 

The parameters were derived by 
fitting the spectra with spectral models: 
different models were used according to the nature of each source, 
as described in subsubsections \ref{sec:spec_snr}, \ref{sec:ana_Xpul}, and 
\ref{sec:spec_rem}. 
For sources detected in multiple observations 
(table \ref{tab:asca-asca}), 
we first fitted the spectrum from each observation separately. 
Except for SMC~X-1, 
the spectral parameters 
($\Gamma$, $kT$, and {$N_{\rm H}$}) 
in each observation were found to be consistent with each other. 
We thus simultaneously fitted all of the spectra 
with parameters linked between the observations, 
in order to obtain more stringent constraints. 
However, the flux was not linked 
in the simultaneous fitting, 
in order to examine the flux variability 
(readers should 
note that there could be $\lesssim 10$--20\% error in the flux 
of most sources). 
Hardness ratios for those sources were
derived after adding the spectra from all observations. 

\subsubsection{Hardness ratio\label{sec:hr}}
The spectral hardness ratio (HR) was 
derived by the definition 
${\rm HR} = (H-S)/(H+S)$, 
where $H$ and $S$ represent 
background-subtracted GIS count rates 
in 2.0--7.0~keV and 0.7--2.0~keV, respectively. 
HR is not given for No.\ 27 in table \ref{tab:cat_param} 
because this source was only resolved with SIS (in obs.\ Q). 
For the same reason, 
HR of No.\ 26 was derived only from the data of obs.\ F.

\subsubsection{Spectra of SNRs\label{sec:spec_snr}}
X-rays were detected from the positions of eight radio SNRs\footnote{%
No.\ 83 (AX~J0105$-$722) was once identified with 
SNR DEM~S128 \citep{Yokogawa2000e}, 
but now the identification is questionable 
due to the improved position of the {ASCA} source 
and the high resolution {ROSAT} study \citep{Filipovic2000a}. 
See also subsubsection \ref{sec:3.34s}.}, 
0045$-$734 (No.\ 21), 
0047$-$735 (No.\ 25), 
0057$-$7226 (No.\ 66), 
0102$-$723 (No.\ 81), 
0103$-$726 (No.\ 82), 
0046$-$735 (No.\ 23),     
0049$-$736 (No.\ 36), and 
0056$-$725 (No.\ 64).     
The former five were detected with SIS and 
the latter three were detected only with GIS. 
%
%
%
At first, we investigated the presence of 
emission lines in the spectra with the same method 
described in subsection 3.4 of \citet{Yokogawa2000e}. 
We found evidence of emission lines from 
0045$-$734,
0057$-$7226, 
0102$-$723, 
0103$-$726, and 
0049$-$736,  
and thus we regard these five as thermal SNRs. 
We therefore fitted their spectra with thin-thermal plasma models, 
as described in subsection \ref{sec:snr}. 
For the other SNRs, 
we first fitted the spectra with both a power-law model 
and a thin-thermal model 
in a collisional ionization equilibrium (CIE) state 
\citep{Raymond1977}, 
and finally adopted a power-law for 0056$-$725 
and the CIE thermal model for 0047$-$735 and 0046$-$735, 
for reasons described in subsection \ref{sec:snr}. 
When fitting with thermal models, 
the metal abundances were primarily fixed at 0.2~solar, 
which is the mean value for the SMC ISM 
\citep{Russell1992}, 
unless otherwise mentioned. 
Hereafter, we refer to this abundance value as 
``the SMC abundance.'' 

\subsubsection{Spectra of X-ray pulsars and HMXBs\label{sec:ana_Xpul}}
The spectra of X-ray pulsars 
(regardless of whether they are accretion-powered or rotation-powered)
are generally described by a power-law 
in the {ASCA} bandpass. 
Therefore, 
we adopted a power-law model for the 
22 detected pulsars 
(summarized in table \ref{tab:pulsars}) 
and also for No.\ 63 (a Be/X-ray binary, RX~J0058.2$-$7231).

Several sources exhibited systematic deviation from the simple power-law. 
Since No.\ 49 (SMC~X-2) and No.\ 90 (XTE~J0111.2$-$7317) 
showed bump-like residuals around 6--7~keV, 
we added a narrow Gaussian line to the model. 
For No.\ 67 (RX~J0059.2$-$7138), the power-law model 
exceeded the data at $\gtrsim 7$~keV, and thus 
we included a high-energy cutoff in the model. 
The brightest pulsars (RX~J0059.2$-$7138, XTE~J0111.2$-$7317, 
and SMC~X-1) all exhibited large data excess over the power-law 
at $\lesssim 2$~keV, thus we added a blackbody component 
to describe the soft excess. 
Details of the analyses and comments are 
given for each source in subsection \ref{sec:pulsar}.

\subsubsection{Remaining sources\label{sec:spec_rem}}
Although the nature of X-ray emission from the remaining sources 
is unclear at this moment, we basically adopted a power-law model 
in the spectral fitting. 
Since No.\ 22 (AX~J0048.2$-$7309) showed weak evidence for an emission line 
at around 6--7~keV, we added a Gaussian line to the model 
(see subsubsection \ref{sec:no22}). 
For No.\ 2 and No.\ 13, 
no constraint 
on the spectral parameters could be obtained 
due to the highly limited statistics; 
we thus do not present the parameters in 
table \ref{tab:cat_param}. 
For No.\ 39, the best-fit model 
($\Gamma = 10$ and $N_{\rm H} = 1.4 \times 10^{23}$~cm$^{-2}$) 
yielded a very high luminosity of $L_{\rm X} \sim 3 \times 10^{39}$~erg~s$^{-1}$. 
Such a high luminosity is unrealistic and is probably 
an artifact caused by the large $\Gamma$ and {$N_{\rm H}$}; 
we thus do not present {$L_{\rm X}$}\ in table \ref{tab:cat_param}. 
For very soft sources (No.\ 6 and No.\ 45), 
we present the results from both a power-law model 
and a CIE thermal model. 

\section{Comments on Specific Sources}
\subsection{X-Ray Pulsars\label{sec:pulsar}}
\begin{figure}
 \FigureFile(80mm,80mm){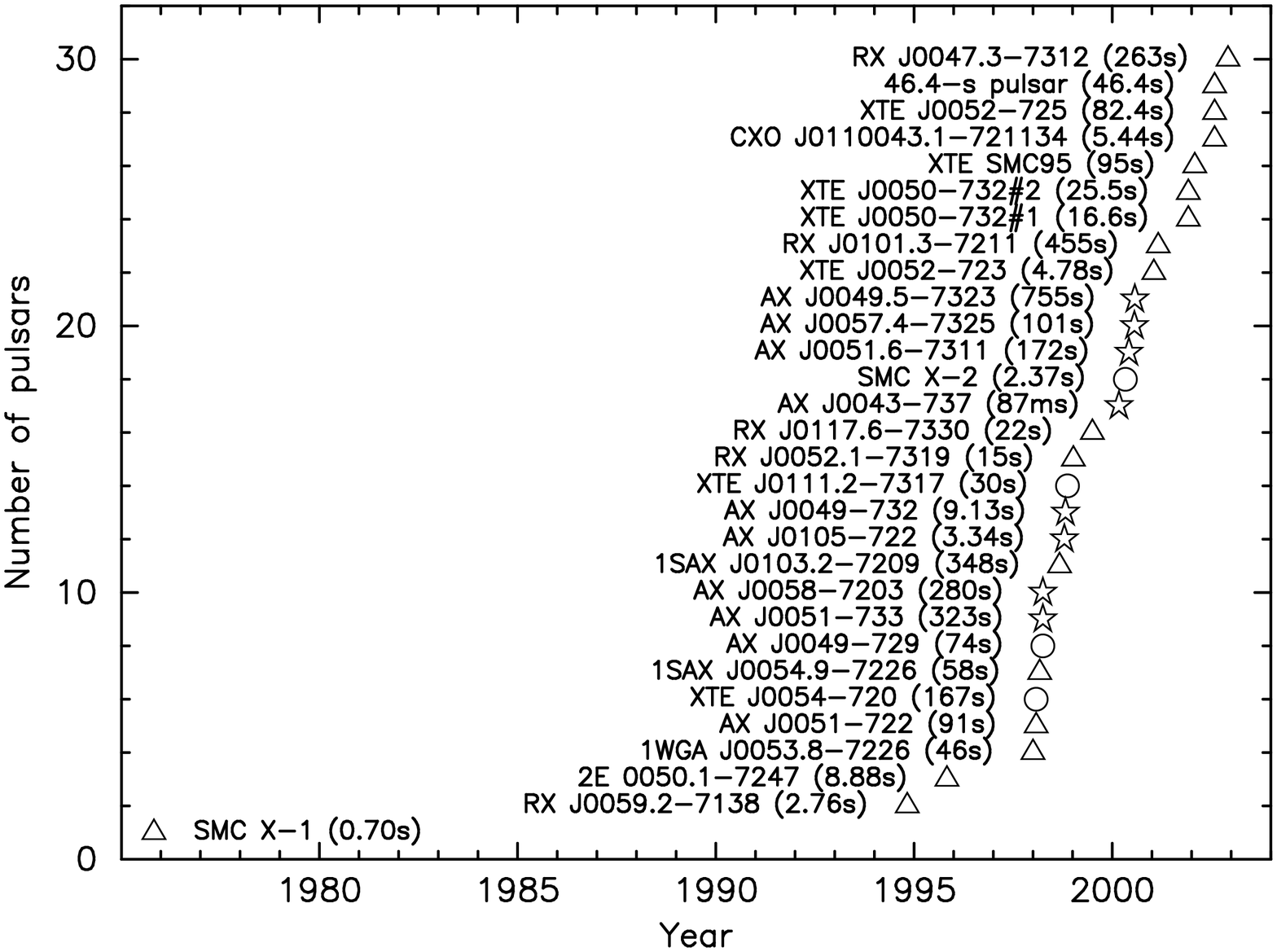}
 \caption{History of discoveries of X-ray pulsars in the SMC 
as of 2002 November. 
The symbols represent 
the pulsars discovered in this study (stars), 
the pulsars for which only the positions were determined 
in this study (circles), 
and the remainder (triangles). 
\label{fig:pulsar_dis}}
\end{figure}

Since the first X-ray pulsar in the SMC, SMC~X-1, 
was discovered \citep{Lucke1976}, 
only three pulsars had been known for about 20 years 
\citep{Hughes1994b,Israel1997}. 
In the last four years, however, there has been 
a rush of pulsar discoveries (see figure \ref{fig:pulsar_dis}), 
and now there are 30 pulsars known in the SMC (table \ref{tab:pulsars}).

In this subsection, we give brief comments on all of the X-ray pulsars 
in order to summarize their nature. 
Since no new information has been obtained for 
XTE~J0055$-$724, 2E~0050.1$-$7247, and RX~J0117.6$-$7330, 
we give the same comments as described in \citet{Yokogawa2000e}. 
%
The pulse periods are designated with 
the error for the last digit in parentheses. 
We regarded a pulsar as an XBP 
if the pulsar had a long pulse period ($\sim 1$--1000~s), 
hard spectrum ($\Gamma \sim 1$), 
flux variability, and/or an optical counterpart. 
Finally we classified 26 out of the 30 pulsars as XBPs. 
%
The spectra and pulse shapes are given in the references for each pulsar, 
and are also summarized in \citet{Yokogawa2002b}.

\subsubsection{No.\ 17 --- AX~J0043$-$737\label{sec:87ms}}

Coherent pulsations of a 87.58073(4)~ms period from AX~J0043$-$737 
were discovered by 
Yokogawa and Koyama (2000) 
in obs.\ K, 
using a timing resolution of 31.25~ms 
(1/16 of the nominal value) as described in subsection \ref{sec:timing_ana}. 
The significance of the detection is at a marginal level, 
$\sim 99.98$\%. 
AX~J0043$-$737 was also detected in a follow-up observation 
with a longer exposure time (obs.\ P). 
Although we performed the FFT on the events in the same energy band, 
no significant peak was found. 
The count rate, total count (without background), and background level 
in 0.5--7.7~keV are 
$2.5 \times 10^{-3}$~count~s$^{-1}$, 200~count, and 64\% in obs.\ K, 
and 
$1.0 \times 10^{-3}$~count~s$^{-1}$, 186~count, and 84\% in obs.\ P. 
%
%
Although the total count is nearly identical in these two observations, 
the smaller count rate and larger background level in obs.\ P 
may have caused the pulsations to be hidden in the background. 
Therefore, a confirmation of the pulsations by 
observations with much higher S/N ratios 
is still needed. 

The spectral shape ($\Gamma$ and {$N_{\rm H}$}) is consistent between the 
two observations, 
while {$F_{\rm X}$}\ shows a slight decline (table \ref{tab:cat_param}). 
The photon index ($\sim 1.7$) is softer than those for usual XBPs, 
$\Gamma \sim 1$ (e.g., Nagase 1989). 
A short pulse period, soft spectrum, and 
luminosity far smaller than the Eddington limit for a neutron star 
are also detected from SAX~J0635$+$0533 ($P=33.8$~ms and $\Gamma = 1.50$), 
which has a Be star counterpart 
\citep{Kaaret1999,Cusumano2000}.
%
%
\citet{Cusumano2000} argued that SAX~J0635$+$0533 
may be an accretion-powered Be/X-ray binary, 
and if so, the magnetic field should be weaker than $2 \times 10^9$~G 
for the accretion to occur against the centrifugal force 
at the magnetosphere. 
%

Since no optical counterpart has been reported for AX~J0043$-$737, 
we investigated the catalogues of emission-line objects 
(potential candidates for Be stars) by 
Meyssonnier and Azzopardi (1993)
and 
Murphy and Bessell (2000), 
but no counterpart was found. 
Therefore, a search for the optical counterpart is needed, 
in addition to confirmation of the 87-ms pulsations.

\subsubsection{No.\ 20 --- RX~J0047.3$-$7312 = IKT1\label{sec:IKT1}}

We tentatively consider No.\ 20, 
RX~J0047.3$-$7312, and IKT1 to be identical 
simply because of the positional coincidence. 
Haberl and Sasaki (2000) 
proposed RX~J0047.3$-$7312 as a Be/X-ray binary candidate 
because this source exhibits a flux variation with a factor of 9 
and has an emission-line object as a counterpart. 
The GIS spectra in two observations (F and Q) are hard, 
having a photon index of $\sim 1$, 
which is typical of Be/X-ray binaries. 
In addition, a flare-like behaviour was found 
from the light curve in obs.\ Q (figure \ref{fig:lc_IKT1}). 

Archival data from an {XMM-Newton} observation 
revealed that 
this source is pulsating with a period of $263(1)$~s 
(M.\ Ueno, private communication). 
All of these results indicate that 
RX~J0047.3$-$7312 is probably an XBP with a Be star companion 
(hereafter, a Be-XBP).

\subsubsection{No.\ 24 --- AX~J0049$-$729}
\citet{Corbet1998}, 
using {RXTE}, first discovered pulsations 
with a 74.8(4)~s period in the direction of SMC X-3, 
with a large positional uncertainty of $\sim 2^\circ$. 
During this study, 
we found that AX~J0049$-$729 was pulsating with a 74.68(2)~s period 
in obs.\ F \citep{Yokogawa1999}. 
The position was determined more accurately and 
is consistent with the 74.8-s pulsar discovered with {RXTE}. 
The {ROSAT} counterpart of this pulsar, RX~J0049.1$-$7250,
provides the most accurate position with a $\pm 13''$ error circle
\citep{Kahabka1998}, 
in which one Be star has been discovered 
\citep{Stevens1999}.
\citet{Yokogawa1999} found a large flux variability with a factor of 
$\gtrsim 100$ 
using archival data of {Einstein} and {ROSAT}. 
From all of the information, we conclude that 
AX~J0049$-$729 is a Be-XBP. 

\subsubsection{No.\ 26 --- AX~J0049$-$732\label{sec:9s}}
Coherent pulsations with a 9.1320(4)~s period 
from AX~J0049$-$732 
were first detected during this study in the following manner. 
Using the data of obs.\ F, 
we performed an FFT on the events in 1.0--5.1~keV collected from 
a $3'$-radius circle centered on AX~J0049$-$732, 
and discovered the coherent pulsations 
with 99.99\% confidence 
\citep{Imanishi1998,Ueno2000a}. 
However, 
since AX~J0049$-$732 was detected at only $\sim 1\farcm 8$ from 
No.\ 25 (SNR 0047$-$735), 
mutual contamination was not negligible due to the 
poor resolution of the {ASCA} XRT. 
Therefore, we also performed an FFT on the events 
in a $3'$-radius circle centered on SNR 0047$-$735, 
and detected no pulsations. 
These analyses clearly indicate that 
the 9.13-s pulsations are attributable to AX~J0049$-$732. 

AX~J0049$-$732 was also detected in obs.\ Q. 
However, in this observation 
another source, No.\ 27, was detected at $\sim 1\farcm 3$ 
from AX~J0049$-$732, 
in addition to SNR 0047$-$735 at $\sim 1\farcm 8$ apart. 
The SIS data were used to resolve these sources 
and determine the spectral parameters, 
but were not used for an FFT analysis 
because the timing resolution (8~s) is insufficient 
to detect the $\sim 9$-s pulsations. 
We therefore used the GIS data for an FFT analysis, 
although No.\ 27 and AX~J0049$-$732 were not resolved. 
We collected photons from 
a $3'$-radius circle centered on AX~J0049$-$732 
and performed the FFT analysis, but 
no significant pulsations were detected. 
This was probably due to 
the poor statistics caused by 
the reduced flux of AX~J0049$-$732 and large contamination 
from No.\ 27, which has 
the same flux level as AX~J0049$-$732 
(see table \ref{tab:cat_param}). 
Therefore, 
a follow-up observation with better spatial resolution 
is needed to confirm the pulsations from AX~J0049$-$732.

Two {ROSAT} sources, 
RX~J0049.2$-$7311 and RX~J0049.5$-$7310 
(No.\ 430 and No.\ 427 in Haberl et al.\ 2000, respectively), 
were found near AX~J0049$-$732 \citep{Filipovic2000a}. 
Since RX~J0049.2$-$7311 has an emission-line object as a counterpart, 
while  RX~J0049.5$-$7310 does not, 
\citet{Filipovic2000a} suggested that 
RX~J0049.2$-$7311 is likely to be the counterpart for AX~J0049$-$732. 
However, according to the updated position of the {ASCA} source, 
the separation between  AX~J0049$-$732 and RX~J0049.2$-$7311 
is $\sim 80''$, which is much larger than the improved positional accuracy 
for {ASCA} sources (see subsubsection \ref{sec:absolute}).  
Therefore, the identification made by 
\citet{Filipovic2000a} is now questionable. 
On the other hand, the separation between 
RX~J0049.5$-$7310 and AX~J0049$-$732 
is $\sim 33''$, thus we regard this source as the counterpart.

\subsubsection{No.\ 30 --- AX~J0049.5$-$7323\label{sec:755s}}
Coherent pulsations with a 755.5(6)~s period from AX~J0049.5$-$7323 
were first discovered during this study 
from the data of obs.\ Q 
\citep{Ueno2000b,Yokogawa2000a}. 
%
%
%
\citet{Yokogawa2000a} investigated 
the archival data of 19 observations of {ROSAT} and {Einstein} 
in which AX~J0049.5$-$7323 was covered, 
and found 
a flux variation with a factor of at least $\sim 10$. 
All of the information indicates that 
AX~J0049.5$-$7323 is an XBP. 
In addition, 
the {ROSAT} counterpart, RX~J0049.7$-$7323, 
is proposed to be a Be/X-ray binary 
because an emission-line object exists in the error circle 
of $\sim 15''$-radius \citep{Haberl2000b}. 
An optical spectroscopy of the emission-line object 
is encouraged to examine whether it is a Be star or not.

\subsubsection{No.\ 32 --- AX~J0051$-$733}
Coherent pulsations with a 323.2(5)~s period from AX~J0051$-$733 
were first discovered during this study from the data of obs.\ F 
and reported by 
Yokogawa and Koyama (1998a) 
and \citet{Imanishi1999}. 
New results were obtained from a long follow-up observation (obs.\ Q): 
AX~J0051$-$733 was detected at a larger flux 
(see table \ref{tab:cat_param}) 
and the pulsations were detected again 
with a shorter period of 321.0(1)~s. 
%
%
%
\citet{Imanishi1999} investigated the archival data of 16 observations 
of {ROSAT} and {Einstein} in which AX~J0051$-$733 was covered, 
and found a flux variability of a factor $\gtrsim 10$. 
In addition, a gradual flux increase 
with a factor of $\sim 2$ was found during obs.\ Q. 
%
The {ROSAT} counterpart, RX~J0050.8$-$7316, 
has a Be star in its error circle \citep{Cowley1997}. 
All of this information indicates that 
AX~J0051$-$733 is a Be-XBP. 

From the empirical relation between 
the pulse and orbital periods of Be-XBPs \citep{Corbet1984}, 
the orbital period of AX~J0051$-$733 is predicted to be $\sim 185$~d. 
On the other hand, 
due to an optical photometric study of the Be star counterpart, 
strong evidence for a modulation with a 0.7-d period was found 
\citep{Coe2000b,Coe2002}. 
Coe and Orosz (2000) 
argued that 
if the modulation is due to the orbital motion, 
the orbital period should be 1.4~d. 
The large discrepancy between Corbet's empirical law 
and the period seen in the optical band is rather problematic, 
and must be solved by future studies.

%

\subsubsection{No.\ 37 --- AX~J0051$-$722\label{sec:92s}}
Coherent pulsations with a $\sim92$~s period were first discovered in
the direction of SMC~X-3 during an {RXTE} observation on 
1997 November 15
\citep{Marshall1997}. A TOO (Target Of
Opportunity) observation with {ASCA} on December 12 (obs.\ H) revealed
that there were two new pulsars, AX~J0051$-$722 and 
1WGA~J0053.8$-$7226, with periods of 91.12(5)~s and 46.63(4)~s,
respectively \citep{Corbet1998}.
At first, AX~J0051$-$722 steadily faded after its discovery, 
and then was found to rebrighten 
at the ends of March and July in 1998 
\citep{Lochner1998a,Lochner1998b,Israel1998a}.  
The spacing of these two flares and 
the initial outburst on 1997 November is $\sim 120$~d. 


\citet{Stevens1999} carried out optical spectroscopic observations 
and discovered a Be star counterpart for AX~J0051$-$722. 
Thus, AX~J0051$-$722 is classified as a Be-XBP. 
Therefore, the $\sim 120$-d spacing of X-ray flares 
could be interpreted as the orbital period of the neutron star 
in this system \citep{Israel1998a}, 
which is supported by the empirical law 
between the pulse and orbital periods \citep{Corbet1984}.


\subsubsection{No.\ 40 --- AX~J0051.6$-$7311}\label{sec:172s}
Coherent pulsations with a 172.40(3)~s period from AX~J0051.6$-$7311 
were first discovered during this study 
from the data of obs.\ Q 
\citep{Torii2000a,Yokogawa2000b}. 
%
%
%
AX~J0051.6$-$7311 has been covered by 
17 observations of {ROSAT} and {Einstein}. 
\citet{Yokogawa2000b} investigated these archival data 
and found a flux variation with a factor of $\gtrsim 20$. 
In addition, the {ROSAT} counterpart, RX~J0051.9$-$7311, 
has been identified with a Be star \citep{Cowley1997}. 
Therefore, 
all of the information 
indicates that AX~J0051.6$-$7311 is a Be-XBP.

\subsubsection{No.\ 43 --- RX~J0052.1$-$7319}

Coherent pulsations with a 15.3~s period from RX J0052.1$-$7319
were discovered in 
contemporaneous {ROSAT} and {CGRO} observations in 1996
\citep{Lamb1999}. 
\citet{Kahabka2000} investigated the data of two {ROSAT} HRI observations 
in 1995 and 1996 
and found a large change of flux with a factor of $\sim 200$. 
The unabsorbed luminosities in the {ROSAT} band were determined to be 
$2.6 \times 10^{35}$~erg~s$^{-1}$\ and 
$5.2 \times 10^{37}$~erg~s$^{-1}$ (in 0.1--2.4~keV; P.\ Kahabka 2001, 
private communication), 
assuming a photon index of 1.0 and a column density of 
$3 \times 10^{21}$~cm$^{-2}$. 
\citet{Israel1999} searched for an optical counterpart 
in the $10''$ error circle and discovered 
a Be star, in addition to a fainter object with an unknown spectral type. 
From all of the information, 
RX~J0052.1$-$7319 is thought to be a Be-XBP.

A faint {ASCA} source, No.\ 43, is positionally coincident 
with RX~J0052.1$-$7319. 
The {ASCA} spectrum puts almost no constraint on the parameters 
(see table \ref{tab:cat_param})
and is consistent with $\Gamma$ and {$N_{\rm H}$}\ assumed by \citet{Kahabka2000}. 
We detected no sign of coherent pulsations 
from either an FFT analysis or an epoch folding search, 
which is probably due to the highly limited statistics. 
Therefore, it is not clear whether No.\ 43 is really
the counterpart of RX~J0052.1$-$7319.

\subsubsection{No.\ 44 --- XTE~J0054$-$720} 

A transient pulsar XTE~J0054$-$720, 
with a period of $\sim 169$~s, 
was discovered with {RXTE} \citep{Lochner1998a}. 
A flux variation and a monotonous spin-up were found 
in the initial {RXTE} observations. 
%
%
Since the {RXTE} error circle was rather large ($10'$ radius), 
identification with other sources has been difficult. 
In fact, five {ROSAT} HRI sources \citep{Sasaki2000} 
are located within the error circle. 
During this study, we detected coherent pulsations with a 167.8(2)~s period 
from No.\ 44 (AX~J0052.9$-$7157) 
and determined its position accurately \citep{Yokogawa2001a}. 
We found that AX~J0052.9$-$7157 is located within the 
error circle of XTE~J0054$-$720 and 
has a variable Be/X-ray binary, RX~J0052.9$-$7158 \citep{Cowley1997}, 
as a counterpart. 
From the nearly equal pulse period and the positional coincidence, 
we conclude that the {ASCA}, {ROSAT}, and {RXTE} sources are identical, 
and thus XTE~J0054$-$720 is a Be-XBP.

\subsubsection{No.\ 47 --- 1WGA~J0053.8$-$7226 = XTE~J0053$-$724}
Coherent pulsations from 1WGA~J0053.8$-$7226 
with a 46.63(4)~s period were discovered in obs.\ H, as described in 
subsubsection \ref{sec:92s} \citep{Corbet1998}.
\citet{Buckley2001} investigated the archival data of 21 {ROSAT} observations, 
and found a large flux variability. 
They also carried out follow-up optical and infrared observations, 
and discovered two Be stars in the error circle. 
All of the information indicates that 
1WGA~J0053.8$-$7226 is a Be-XBP.

\subsubsection{No.\ 49 --- SMC~X-2}
SMC~X-2 is a long-known transient Be/X-ray binary 
with a maximum luminosity of $\sim 10^{38}$~erg~s$^{-1}$\ 
\citep{Clark1978,Murdin1979}. 
%
Large outbursts have been detected 
with SAS-3, HEAO1, and {ROSAT} 
\citep{Clark1978,Marshall1979,Kahabka1996}.

\citet{Corbet2001b} detected a large outburst
($\sim 10^{38}$~erg~s$^{-1}$) 
in the direction of SMC~X-2 in 2000 January--April, 
with the All-Sky Monitor onboard {RXTE}. 
The position was determined with an error radius of $3'$, 
and SMC~X-2 is located near the edge of the error circle. 
Coherent pulsations were also discovered during the outburst, 
with periods of 
2.371532(2)~s on April 12
and 
2.371861(3)~s on April 22--23. 
%
We made a follow-up {ASCA} observation (obs.\ R) 
and detected No.\ 49 at the position 
consistent both with SMC~X-2 and the 2.37-s pulsar. 
We also detected pulsations with a 2.37230(4)~s period 
\citep{Yokogawa2001b}, 
which is in full agreement with the {RXTE} result, 
indicating that the {RXTE} pulsar and No.\ 49 are identical. 
%

The emission line seen in the spectrum (subsubsection \ref{sec:ana_Xpul}) 
has a center energy of 6.3 (6.1--6.5)~keV, 
which is consistent with the K-shell emission from 
neutral or low-ionized Fe, 
and has an equivalent width of 400 (150--640)~eV. 
%
%
\citet{Yokogawa2001b} carried out a pulse-phase-resolved 
spectroscopy, and found marginal evidence for pulsations of the Fe line 
intensity.

\subsubsection{No.\ 51 --- XTE~J0055$-$724 = 1SAX~J0054.9$-$7226} 
A scan observation made with {RXTE} on 1998 January 20 
revealed a new X-ray pulsar, XTE~J0055$-$724, 
with a pulse period of $\sim 59$~s \citep{Marshall1998}.
\citet{Santangelo1998} made a follow-up observation 
with {SAX} on January 28, 
and detected pulsations with a 58.963(3)~s period 
from 1SAX~J0054.9$-$7226, which is located within 
the $10'$ error circle of XTE~J0055$-$724. 
The agreement of the period and position indicates 
that XTE~J0055$-$724 and 1SAX~J0054.9$-$7226 are the same source. 
During this study, we detected weak evidence for $\sim 59$~s pulsations 
from No.\ 51, 
which is positionally coincident with 1SAX~J0054.9$-$7226; 
hence, we consider No.\ 51 to be 
the counterpart for this pulsar. 

\citet{Israel1998b} investigated the archival data of 13 {ROSAT} observations 
covering 1SAX~J0054.9$-$7226. 
They found a flux variation with a factor of $> 30$ 
between two observations of {ROSAT} in 1996 and {RXTE} in 1998. 
They also determined the position with a $10''$ error circle, 
in which a Be star was later discovered by 
\citet{Stevens1999}.
XTE~J0055$-$724 = 1SAX~J0054.9$-$7226 
is thus a Be-XBP.

\subsubsection{No.\ 56 --- AX~J0057.4$-$7325}
Coherent pulsations with a 101.45(7)~s period from AX~J0057.4$-$7325 
were first discovered during this study 
from the data of obs.\ R 
\citep{Torii2000b,Yokogawa2000c}. 
This source was also found in obs.\ L, 
and weak evidence for pulsations with a period of 101.47(6)~s was detected. 
\citet{Yokogawa2000c} investigated six {ROSAT} observations 
covering this pulsar, and found a flux variability 
with a factor of $> 10$. 
All of the information indicates that 
AX~J0057.4$-$7325 is an XBP.

So far optical follow-up observations have not been carried out. 
As far as we have investigated, 
only one optical source, MACS~J0057$-$734\#010, is located within 
the {ASCA} error circle \citep{Tucholke1996}, 
for which the spectral type and existence of the H$\alpha$ emission line 
are not known. 
No counterpart is found in the catalogues of
emission-line objects by 
Meyssonnier and Azzopardi (1993)
and 
Murphy and Bessell (2000). 
This is a rare case in which an XBP in the SMC is not associated 
with a Be star or an emission-line object (see Haberl, Sasaki 2000). 
We note that AX~J0057.4$-$7325 
is located at the edge of the SMC main body, 
fronting the eastern wing. 
The fact that OB supergiant X-ray binaries 
(only SMC~X-1 and EXO~0114.6$-$7361; see tables \ref{tab:pulsars} and \ref{tab:npHMXBs}) 
are both located in the eastern wing 
leads us to suspect that AX~J0057.4$-$7325 
might be the third example. 
Therefore, deep and detailed optical observations around this pulsar 
are strongly encouraged.

\subsubsection{No.\ 61 --- AX~J0058$-$7203}

Coherent pulsations with a period of 280.4(4)~s from 
AX~J0058$-$7203 were first discovered during 
this study \citep{Yokogawa1998a,Tsujimoto1999}. 
%
In a new observation T, AX~J0058$-$7203 was detected at 
a large off-axis angle of $\sim 23'$. 
We performed an FFT analysis 
and found no significant pulsations. 
%
The count rate, total counts (without background), and background level 
in 1.0--6.0~keV are 
$1.3 \times 10^{-2}$~count~s$^{-1}$, 1014~count, and 39\% in obs.\ G, 
and 
$7.3 \times 10^{-3}$~count~s$^{-1}$,  642~count, and 42\% in obs.\ T. 
%
%
The reduced total counts may cause the non detection of pulsations, 
although no quantitative estimation has been done.

\citet{Tsujimoto1999} investigated 
archival data of 14 observations with {Einstein} and {ROSAT} 
covering this pulsar, 
and found a flux variation with a factor of $\gtrsim 10$. 
%
Haberl and Sasaki (2000) found an emission-line object 
as a counterpart for AX~J0058$-$7203. 
All of this information indicates that AX~J0058$-$7203 is an XBP, 
and probably has a Be star companion. 
%

\subsubsection{No.\ 67 --- RX~J0059.2$-$7138\label{sec:2.7s}}
A transient source, RX~J0059.2$-$7138, was discovered 
with {ROSAT} and {ASCA} 
in simultaneous observations of SNR 0102$-$723 (obs.\ B). 
Coherent pulsations with a period of 2.7632(2)~s were discovered 
from the {ROSAT} data \citep{Hughes1994b}, 
and were confirmed with the {ASCA} data \citep{Kohno2000}. 
The pulsed fraction is 
larger at higher energy \citep{Hughes1994b,Kohno2000}. 
The possible optical counterpart proposed by \citet{Hughes1994b} 
was later revealed to be a Be star \citep{Southwell1996}, 
and thus RX~J0059.2$-$7138 is undoubtedly a Be-XBP.

The {ASCA} spectrum exhibits a soft excess below $\lesssim 2$~keV 
and a cut-off above $\gtrsim 7$~keV, 
as described in subsubsection \ref{sec:ana_Xpul}. 
We thus included a blackbody model as the soft component 
and a high-energy exponential cut-off. 
\citet{Kohno2000} used 
the {ROSAT} and {ASCA} data simultaneously, and carried out 
more elaborate analyses. 
They adopted various models to describe the soft excess, 
and found that the soft component could be described 
by a thin-thermal plasma model or a broken power-law 
combined with an oxygen overabundance in the absorption column. 
They also found that 
the normalization of the soft component does not change 
(i.e., the soft component exhibits no pulsations) 
during the pulse phase, 
which is consistent with the energy-dependent pulsed fraction. 

\subsubsection{No.\ 72 --- CXOU~J0110043.1$-$721134}
\citet{Lamb2002} 
carried out a 100-ks {Chandra} observation, 
and discovered coherent pulsations 
with a period of 5.44~s from CXOU~J0110043.1$-$721134. 
Its spectrum was 
fitted by a blackbody model with a temperature of $\sim 0.41$~keV 
and a luminosity of $\sim 10^{35}$~erg~s$^{-1}$. 
They investigated the archival data from other satellites, 
and found little flux variation. 
From all of the information, 
\citet{Lamb2002} proposed that 
CXOU~J0110043.1$-$721134 is an anomalous X-ray pulsar (AXP). 

We found that 
the ASCA spectrum of No.\ 72, originally 
fitted by a soft power-law model ($\Gamma \sim 3$), 
could also be described 
by a blackbody model with a temperature of $\sim 0.41$~keV. 
This fact confirms that No.\ 72 is the counterpart for 
CXOU~J0110043.1$-$721134.

\subsubsection{No.\ 74 --- RX~J0101.3$-$7211}\label{sec:455s}
RX~J0101.3$-$7211 has been known to be a highly variable {ROSAT} source 
with an emission-line object 
\citep{Haberl2000b}, 
and was classified as an XBP candidate 
in our earlier study (source No.\ 27 in Yokogawa et al.\ 2000e) 
because of its hard spectrum. 
Therefore, this source has been	strongly suspected to be a Be-XBP, 
although no pulsations were detected, 
probably because of the limited statistics.

\citet{Sasaki2001} analyzed an {XMM-Newton} observation 
on SNR 0102$-$723 and serendipitously detected RX~J0101.3$-$7211. 
Although the flux level was the lowest among 
previous detections with {ROSAT} and {ASCA}, 
coherent pulsations with a 455(2)~s period were discovered 
thanks to the high $S/N$ ratio achieved by {XMM-Newton}. 
The pulsed fraction was found to be energy-dependent 
and higher at lower energy, the same as AX~J0049.5$-$7323 
\citep{Ueno2000b}.
%
\citet{Sasaki2001} also carried out a spectroscopic optical 
observation on the emission-line object found by 
Haberl and Sasaki (2000), 
and revealed it to be a Be star. 
Therefore, RX~J0101.3$-$7211 is undoubtedly a Be-XBP.

\subsubsection{No.\ 78 --- 1SAX~J0103.2$-$7209}

Hughes and Smith (1994) 
and \citet{Ye1995} 
performed {ROSAT} HRI observations of the shell-like radio SNR 
0101$-$724 and detected no X-rays from the radio shell. 
Instead, an X-ray point source RX~J0103.2$-$7209 
(= 1SAX~J0103.2$-$7209) 
having a Be star counterpart 
was detected inside the SNR.

Coherent pulsations with a period of 345.2(3)~s from 1SAX~J0103.2$-$7209 
were first discovered  
in a {SAX} observation made on 1998 July 26--27 
(Israel et al.\ 1998a, 2000). 
Subsequently, pulsations with a period of 348.9(3)~s were 
detected from the {ASCA} source No.\ 78 
on 1996 May 21--23, 
(obs.\ D; Yokogawa, Koyama 1998c). 
\citet{Israel2000} also detected 343.5(5)~s pulsations 
from {Chandra} data obtained on 1999 August 23, 
and found a monotonous spin-up 
with a period derivative of $-1.7$~s~yr$^{-1}$.

1SAX~J0103.2$-$7209 has been detected in 
various observations with {Einstein}, {ROSAT}, {ASCA}, and {Chandra}, 
with nearly the same luminosity level of $\lesssim 10^{36}$~erg~s$^{-1}$\ 
(Israel et al.\ 2000, and references therein). 
According to this fact, 
\citet{Israel2000} argued that this pulsar can possibly be classified as 
a ``persistent'' Be-XBP, such as 
X~Per, RX~J0146.9$+$6121, RX~J0440.9$+$4431, 
and RX~J1037.5$-$564 
\citep{White1983,Mereghetti2000,Reig1999}. 
The 755-s pulsar AX~J0049.5$-$7323 (subsubsection \ref{sec:755s}) 
may also belong to this class, although the upper limit of the 
flux variation is not known.

\subsubsection{No.\ 83 --- AX~J0105$-$722\label{sec:3.34s}}

\begin{figure}
 \FigureFile(53mm,50mm){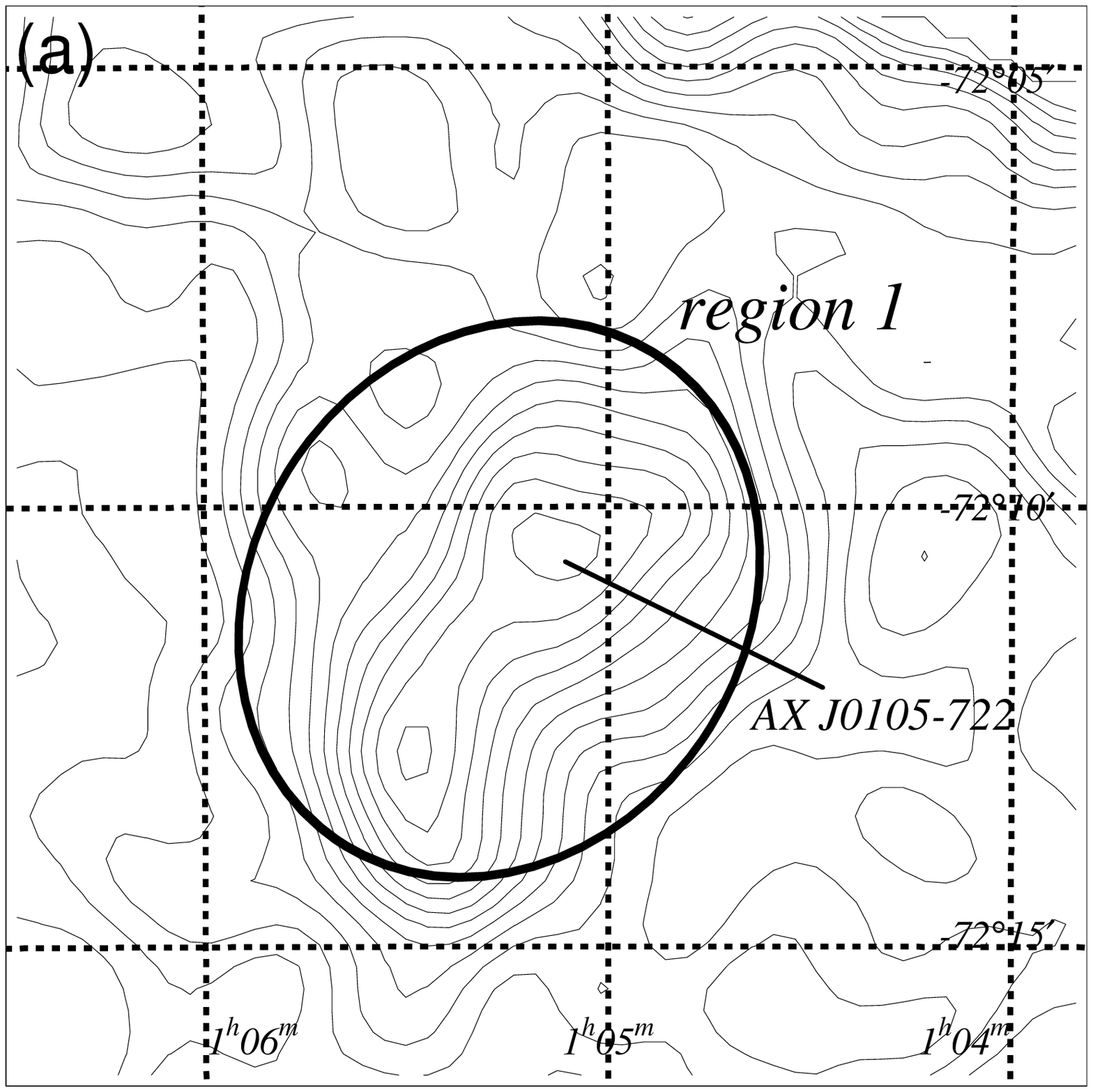}%
 \FigureFile(53mm,50mm){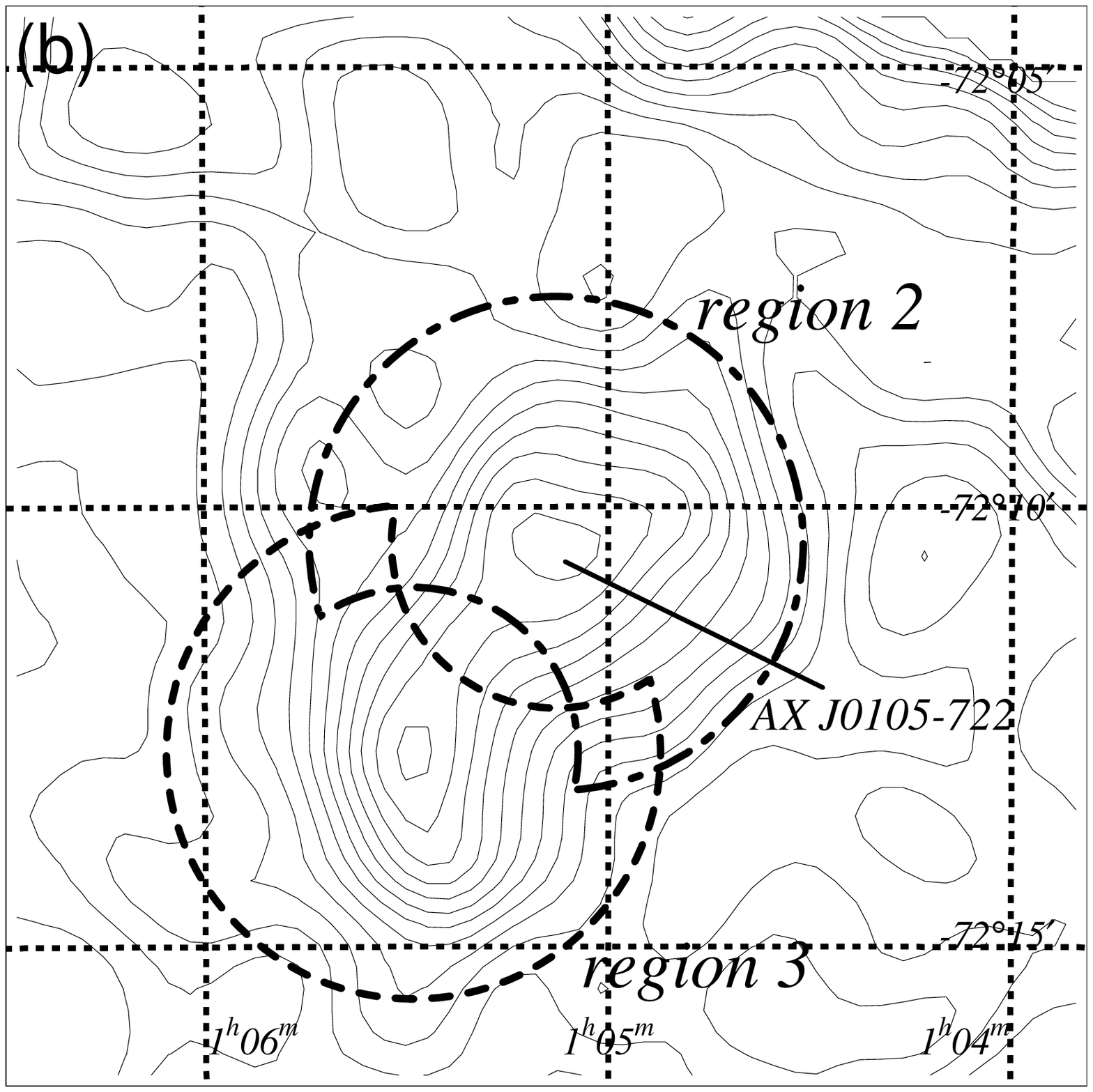}%
 \FigureFile(53mm,50mm){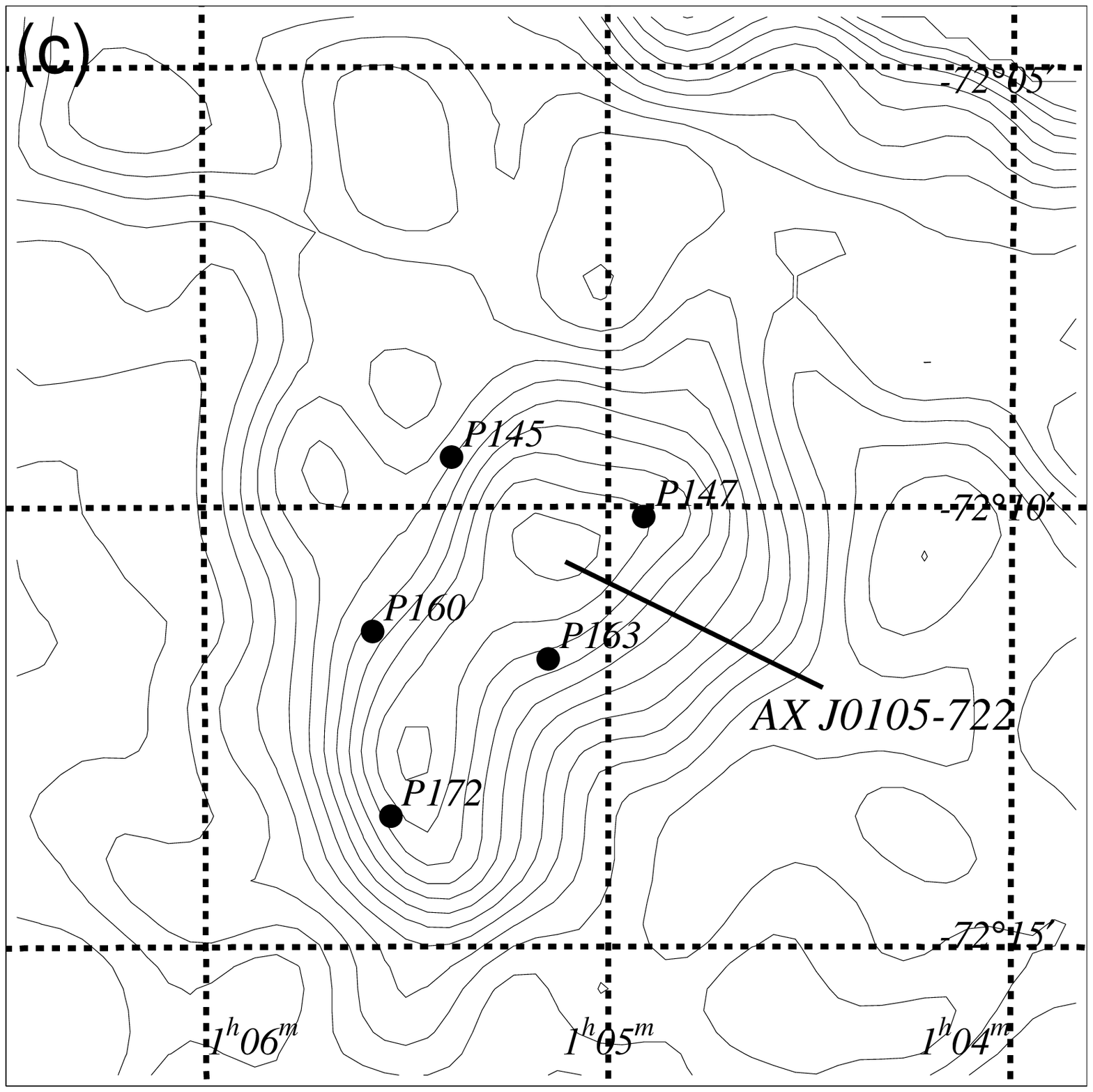}
 \caption{{ASCA} GIS contour images around AX~J0105$-$722 in 0.7--7.0~keV, 
overlaid with the equatorial coordinates (J2000). 
Two {ASCA} sources are detected in this image, one of which 
is AX~J0105$-$722 and the other is No.\ 84, located 
at $\sim 3'$ southeast. 
Regions 1--3 from which event lists were extracted for 
pulsation searches (see text) are indicated by the ellipse 
and circles in (a) and (b). 
The five dots marked with ``P$n$'' in (c) are the {ROSAT} PSPC sources 
in \citet{Haberl2000a} with an ID number of $n$. 
\label{fig:search3.34s}}
\end{figure}

\begin{figure}
 \FigureFile(80mm,80mm){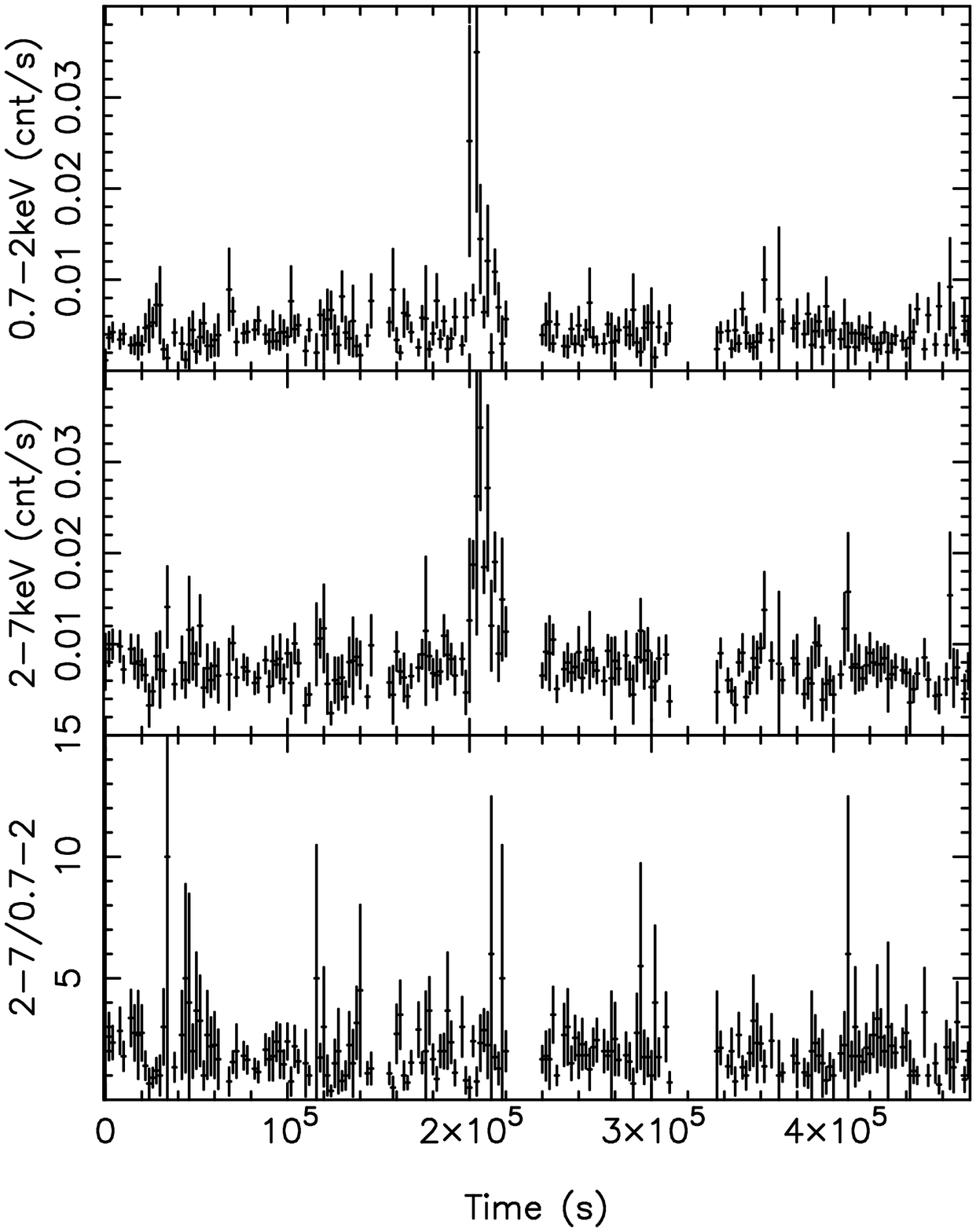}
 \FigureFile(80mm,80mm){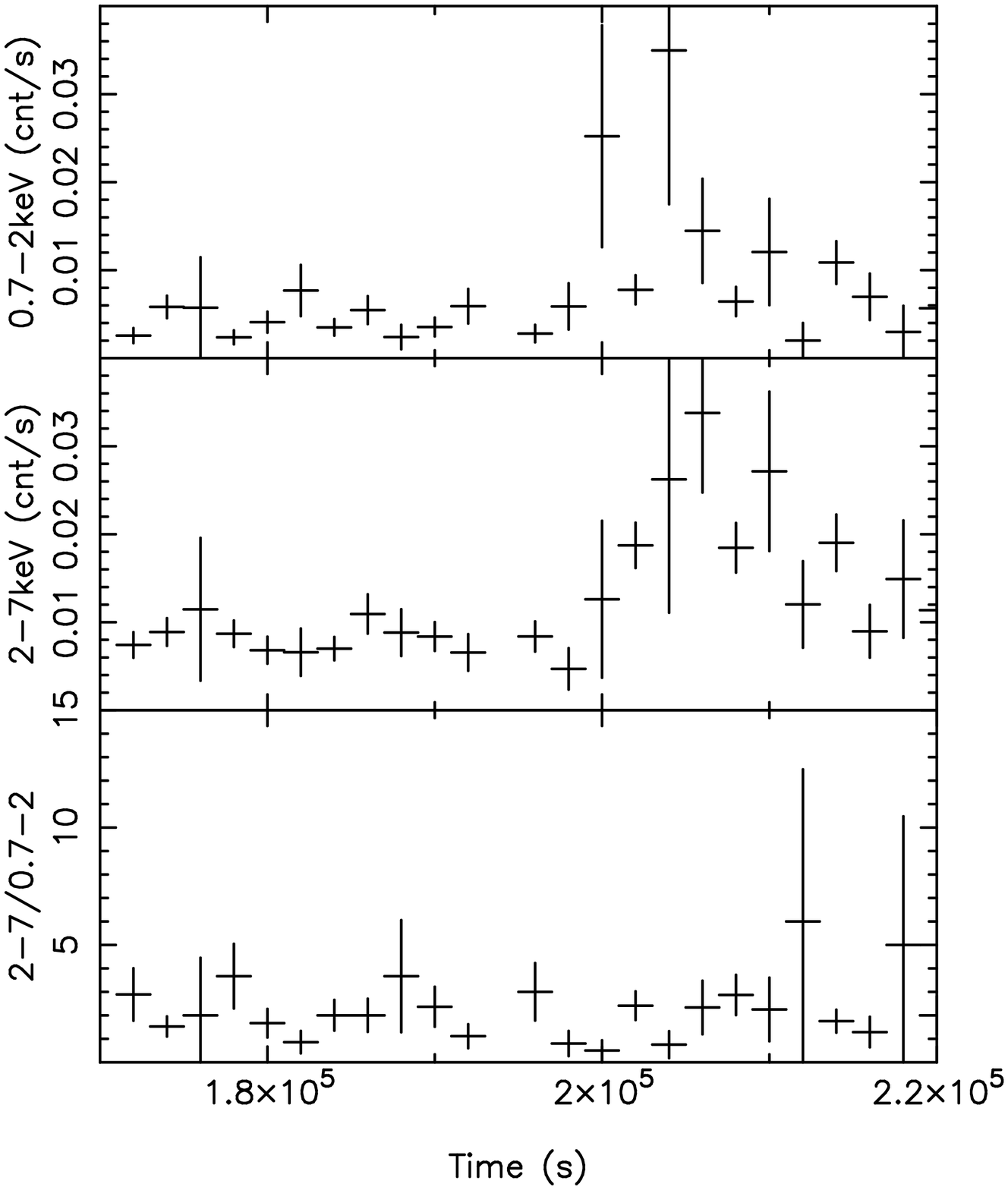}
 \caption{
Light curves of RX~J0047.3$-$7312 (IKT1) in obs.\ Q with a bin time of 2000~s. 
The upper and middle panels present the light curves 
in the soft (0.7--2.0~keV) and hard (2.0--7.0~keV) bands, 
respectively. 
The lower panels present the ratio of the two bands. 
A flare-like activity is found at the time $\sim 2 \times 10^5$~s; 
the right panel shows the time zone around the flare. 
\label{fig:lc_IKT1}}
\end{figure}

Since AX~J0105$-$722 and No.\ 84 
are located only $\sim 3'$ from each other, 
mutual contamination is not negligible. 
We therefore used an oval-shaped region including 
both AX~J0105$-$722 and No.\ 84 
(region 1 in figure \ref{fig:search3.34s}a)
in the timing analysis, 
and detected coherent pulsations with 
a 3.34300(3)~s period at a marginal significance of $\sim 99.5$\% 
\citep{Yokogawa1998b}.
We then separately searched for pulsations from 
regions 2 and 3 in figure \ref{fig:search3.34s}b and found weak evidence 
for the 3.34-s pulsations only from region 2 
(which includes AX~J0105$-$722).
Therefore, we conclude that the pulsations are attributable 
to AX~J0105$-$722. 
The spectrum of AX~J0105$-$722 is rather soft 
(see table \ref{tab:cat_param}) compared to those of typical XBPs, 
$\Gamma \sim 1$ \citep{Nagase1989}.

\citet{Filipovic2000b} used the data from 
high resolution X-ray and radio observations 
around this source made 
with {ROSAT}, ATCA (Australia Telescope Compact Array), and 
MOST (Molonglo Observatory Synthesis Telescope).
They resolved the {ASCA} sources AX~J0105$-$722 and No.\ 84 
into five {ROSAT} PSPC sources as shown in figure \ref{fig:search3.34s}c: 
AX~J0105$-$722 is surrounded by three {ROSAT} PSPC sources, 
Nos.\ 145, 147, and 163 
(here, PSPC-145, PSPC-147, and PSPC-163)
in \citet{Haberl2000a}. 
PSPC-163 (RX~J0105.1$-$7211) 
exhibits a hard spectrum typical of HMXBs 
and is identified with an emission line object catalogued in 
Meyssonnier and Azzopardi (1993), and 
thus is a likely candidate for a Be/X-ray binary 
(e.g., Haberl, Sasaki 2000). 
PSPC-145 is positionally coincident with 
the radio emission from SNR DEM~S128, 
has a soft X-ray spectrum typical of SNRs, 
and exhibits no flux variation, and is thus likely to be 
an X-ray emitting SNR. 
PSPC-147 has the hardest spectrum among the three, 
although the source nature of the source is unclear. 
Considering these facts, we propose that 
the 3.34-s pulsations can be attributed to 
RX~J0105.1$-$7211, which is a Be-XBP, 
and that 
the mutual contamination of X-rays from 
RX~J0105.1$-$7211, PSPC-145, and PSPC-147 
in the {ASCA} data caused 
the rather soft spectrum of AX~J0105$-$722. 
Because of this situation 
and the rather marginal detection of pulsations, 
follow-up X-ray observations with high spatial resolution 
and good S/N ratio are needed.

\subsubsection{No.\ 90 --- XTE~J0111.2$-$7317\label{sec:31s}}

A transient source, XTE~J0111.2$-$7317, was serendipitously 
found with {RXTE} and {CGRO}, and at the same time 
coherent pulsations with a period of $\sim 31$~s were discovered 
\citep{Chakrabarty1998a,Wilson1998}. 
In the TOO observation with {ASCA} (obs.\ I), 
pulsations with a 30.9497(4)~s period 
were unambiguously detected from No.\ 90. 
Thus No.\ 90 is undoubtedly the counterpart for XTE~J0111.2$-$7317, 
and the position was determined with an accuracy better 
than {RXTE} \citep{Chakrabarty1998b}.
\citet{Israel1999} detected two stars in the error circle; 
the fainter one was revealed to be a Be star and was proposed to be 
the optical counterpart for this pulsar. 
\citet{Coe2000a} carried out a more elaborate optical 
spectroscopy on the optical counterpart, and  
confirmed the Be nature of this star. 
Therefore, XTE~J0111.2$-$7317 is undoubtedly a Be-XBP.

The soft X-ray excess shown in the spectrum 
was well fitted with a blackbody model (subsubsection \ref{sec:ana_Xpul}). 
However, the pulse phase resolved spectroscopy performed by 
\citet{Yokogawa2000d} revealed that 
the blackbody component is pulsating, 
while the emission region of the blackbody is extremely large 
($\sim 800$~km in radius) and so pulsation is impossible. 
Therefore, \citet{Yokogawa2000d} proposed an alternative model, 
the ``inversely broken power-law,'' which is a power-law 
with a larger/smaller photon index below/above a break energy, 
to describe the whole continuum and the pulsations in the low energy band. 
The pulsations of the soft component are in striking contrast 
to the non-pulsating soft component of RX~J0059.2$-$7138 (subsubsection 
\ref{sec:2.7s}), 
although the overall continuum shapes in the {ASCA} band 
resemble each other. 

\subsubsection{No.\ 94 --- SMC~X-1\label{sec:x-1}}
SMC~X-1 
is a well-established XBP with a pulse period of $\sim 0.71$~s, 
having a B-type supergiant companion (e.g., Bildsten et al.\ 1997). 
SMC~X-1 has been detected in three {ASCA} observations. 
The 0.71-s pulsations were detected only from the data 
of obs.\ A and obs.\ I. 
In obs.\ C, it was in a low-state and 
went into eclipse as predicted by the ephemeris 
of \citet{Wojdowski1998}. 
Since the low-state of SMC~X-1 should be caused by 
an occultation of the X-ray emitter, 
probably by the tilted accretion disk 
\citep{Wojdowski1998,Vrtilek2001}, 
the intrinsic luminosity in the low-state should be 
as high as that in the high-state. 
Therefore, 
the small {$F_{\rm X}$}\ and {$L_{\rm X}$}\ derived from the obs.\ C data 
(table \ref{tab:cat_param}) appear to be an underestimate of the true intensity.

The soft excess, which was described by a blackbody component 
(subsubsection \ref{sec:ana_Xpul}), was found to be pulsating 
by a phase-resolved spectroscopy performed on the obs.\ A data 
\citep{Paul2002}. 
Therefore, for the same reason as described in subsubsection \ref{sec:31s}, 
attributing the soft excess to blackbody emission 
is not appropriate. 
\citet{Paul2002} found that 
the inversely broken power-law model 
or two-component power-law model could describe 
both the continuum shape and the pulsating nature of the soft excess.

\subsubsection{XTE~J0050$-$732\#1 and \#2}
\citet{Lamb2001} discovered two new pulsars 
with periods of 16.6~s and 25.5~s 
from archival data of {RXTE}. 
Although they gave no names for these pulsars, 
we tentatively designate them as 
XTE~J0050$-$732\#1 (16.6~s) and 
XTE~J0050$-$732\#2 (25.5~s)
because the {RXTE} observation was centered on 
(00{$^{\rm h}$}50$^{\rm m}$44\fs 64, $-$73$^\circ$16$'$04\farcs 8). 
From the long pulse periods and the period derivative 
(found only for XTE~J0050$-$732\#1), 
we regard both as XBPs, 
although more information (optical counterparts and 
flux variability) is required for a further confirmation.


\subsubsection{2E~0050.1$-$7247} 
Coherent pulsations with a 8.8816(2)~s period
from 2E~0050.1$-$7247 = RX~J0051.8$-$7231 were discovered 
in a {ROSAT} observation by \citet{Israel1997}.
A flux variability with a factor 20 between two {ROSAT} observations and
a Be star in the error circle were found; 
hence, this pulsar is a Be-XBP. 
{ASCA} covered the position of 2E~0050.1$-$7247 in obs.\ H.
We did not detect any positive excess above the background level
from this position. 
The upper limit of its flux is estimated to be
$\sim 1 \times 10^{-13}$erg~s$^{-1}$~cm$^{-2}$\ (0.7--10.0~keV),
assuming a photon index of $\sim 1$.

\subsubsection{XTE~J0052$-$723}
A new transient pulsar XTE~J0052$-$723 
with a period of 4.782(1)~s and a flux of 8~mCrab
was discovered with {RXTE} \citep{Corbet2001a}. 
%
The pulse period indicates that this source is an XBP, 
probably a Be-XBP because of the transient nature.
The position was determined with a $2' \times 1'$ error box, 
and is covered by {ASCA} observation H. 
We detected no X-rays from this position, 
and set the upper limit to be 
$\sim 1 \times 10^{-13}$erg~s$^{-1}$~cm$^{-2}$\ (0.7--10.0~keV),
assuming a photon index of $\sim 1$.

\subsubsection{XTE~J0052$-$725 and the other 46.4-s pulsar}
\citet{Corbet2002} 
reported the discovery of two new transient pulsars 
with periods of 82.4(2)~s and 46.4(1)~s 
from {RXTE} observations. 
Their transient nature and the intensity of $\sim 1.5$~mCrab 
indicate that these are probably XBPs. 
The position of the 82.4-s pulsar, XTE~J0052$-$725, 
was determined with an error box of $\sim 2' \times 8'$. 
We found no ASCA source in the error box. 
The position of the 46.4-s pulsar, 
for which no name was given, 
was not well determined \citep{Corbet2002}.

\subsubsection{XTE~SMC95}
\citet{Laycock2002} carried out monitoring observations 
of the SMC with {RXTE}, and discovered a new X-ray pulsar 
(XTE~SMC95) with a period of 95~s. 
The spectrum was found to be hard 
with $\Gamma \sim 1.5$. 
Multiple observations on the same position 
revealed its transient nature. 
All of the information indicates that XTE~SMC95 is an XBP 
(probably a Be-XBP because of the transient nature). 
Because the uncertainty of the position is very large 
($\sim 2^\circ \times 0\fdg 2$), the source location 
should be determined by future studies.

\subsubsection{RX~J0117.6$-$7330} 
RX~J0117.6$-$7330 was serendipitously discovered
in a {ROSAT} PSPC observation 
\citep{Clark1996}.
The luminosity was $2.3\times10^{37}$erg~s$^{-1}$\ 
between 0.2--2.5~keV at that time
\citep{Clark1997},
and was found to diminish by a factor of over 100 within one year.
\citet{Macomb1999} 
discovered coherent pulsations with a $\sim 22.07$~s period
from the same data,
with the aid of the archival data obtained by BASTE onboard {CGRO}
in the same epoch. 
Strong Balmer emission lines and infrared excess were detected
from the companion star in the error circle 
\citep{Coe1998}, 
indicating that RX~J0117.6$-$7330 is a Be-XBP. 
Although the position of RX~J0117.6$-$7330 was covered in
{ASCA} observations A and C, no X-ray emissions were detected.
It was difficult to estimate the upper limits of the flux
because of the contamination from SMC~X-1, which is 
located only $\sim 5'$ away from RX~J0117.6$-$7330.

\subsection{SNRs\label{sec:snr}} 
We give brief comments on the eight SNRs detected with {ASCA}. 
Since no new information is obtained for 0056$-$725 and 0102$-$723, 
we merely give the same comments as described in \citet{Yokogawa2000e}. 

\subsubsection{No.\ 21 --- 0045$-$734 (N19)\label{sec:0045}}
A {ROSAT} HRI image shows 
that X-ray emission from 0045$-$734 is concentrated 
within the diffuse radio emission \citep{Yokogawa2001c}; 
thus, this SNR would be classified as 
a ``centrally brightened'' SNR defined by \citet{Williams1999}. 
\citet{Yokogawa2001c} analyzed the SIS spectrum 
with NEI (non-equilibrium ionization) plasma models, 
and found overabundances of some elements 
which are consistent with the nucleosynthesis of a type-II SN. 
In spite of the overabundance, 
the plasma age was found to be large, $\gtrsim 3\times 10^4$~yr. 
An evolutionary scenario for such old and overabundant SNRs 
is proposed by \citet{Yokogawa2001c}. 

\subsubsection{No.\ 23 --- 0046$-$735}
Both a power-law model and a thermal CIE model could 
describe the GIS spectrum 
well due to highly limited statistics. 
However, we adopted a thermal model because the derived temperature 
is reasonable for an SNR; 
the thermal nature of this SNR is thus suggested. 

\subsubsection{No.\ 25 --- 0047$-$735}

Source and background regions for 0047$-$735 
were carefully chosen to avoid contamination from 
the nearby pulsar AX~J0049$-$732. 
As a result, the extracted SIS spectrum had rather poor statistics, 
which probably caused a non-detection of emission lines 
(subsubsection \ref{sec:spec_snr}). 
Thus, we also used the GIS spectra (in obs.\ F and Q) 
to compensate for the poor statistics of the SIS spectrum, 
and carried out a simultaneous fitting. 

The spectra were well fitted to both the power-law 
and the CIE thermal model. 
However, we adopt the thermal model because 
the best-fit temperature is reasonable for an SNR. 
This fact indicates that X-ray emission from this SNR 
has a thermal origin.

\subsubsection{No.\ 36 --- 0049$-$736}
In our earlier study \citep{Yokogawa2000e}, 
this SNR was regarded as a thermal SNR because of its soft spectrum, 
although no emission lines were detected 
due to highly limited statistics. 
Data from new observations (L and Q) allowed us to detect 
emission lines (subsubsection \ref{sec:spec_snr}), 
and thus the thermal nature is now established. 
The GIS spectra were well fitted to a CIE thermal model. 
The bump-like residuals are found at $\sim 1.9$~keV, 
which corresponds to the energy of the emission line 
from He-like Si. 
However, 
allowing the Si abundance to be free did not improve the result 
within the statistical error.

\subsubsection{No.\ 64 --- 0056$-$725}
Since fitting the spectrum with a thin thermal plasma model yielded 
an unusually high temperature of $\sim 20$~keV, 
we adopted a power-law model. 
There is a possibility that 
X-rays from 0056$-$725 actually have a non-thermal origin. 

\subsubsection{No.\ 66 --- 0057$-$7226 (N66)}

Using a {ROSAT} HRI image, \citet{Yokogawa2001c} found that 
X-ray emission from 0057$-$7226
is concentrated within the radio shell. 
The SIS spectrum was well fitted with 
an NEI plasma model, showing an overabundance of no element 
\citep{Yokogawa2001c}. 
The plasma age was found to be relatively old, 
$\gtrsim 6\times 10^3$~yr.

\subsubsection{No.\ 81 --- 0102$-$723\label{sec:0102}}
A detailed analysis of the SIS spectrum of SNR 0102$-$723 
was carried out by \citet{Hayashi1994}. 
Strong emission lines from various elements 
were detected, which is consistent with this SNR being young. 
We adopted the same model and determined its flux and luminosity 
(table \ref{tab:cat_param}).

\subsubsection{No.\ 82 --- 0103$-$726\label{sec:0103}}
A {ROSAT} HRI image shows that 
X-rays from 0103$-$726 
exhibit faint emission along the radio shell, 
in addition to more prominent emission concentrated 
at the center of the shell \citep{Yokogawa2001c}. 
The SIS spectrum was well fitted with NEI plasma models. 
As for 0045$-$734 (subsubsection \ref{sec:0045}), 
\citet{Yokogawa2001c} found that 
some elements are overabundant, 
which is consistent with the nucleosynthesis of a type-II SN, 
and that 
the plasma age is large, $\gtrsim 1\times 10^4$~yr.

\subsection{Other Interesting Sources}

\subsubsection{No.\ 22 --- AX~J0048.2$-$7309\label{sec:no22}}
AX~J0048.2$-$7309 was detected in two observations (F and Q) 
and shows a hard spectrum ($\Gamma \sim 1$) and 
a flux variability with a factor of $\sim 5$ (table \ref{tab:cat_param}). 
In addition, we found an emission-line object, 
No.\ 215 in 
Meyssonnier and Azzopardi (1993), 
in the error circle of AX~J0048.2$-$7309. 
All the information suggests that 
this source is probably a Be-XBP. 
Follow-up observations for pulsation searches 
and optical identification are thus encouraged. 

Although the GIS spectra were well fitted to a simple power-law, 
there remains a bump-like residual in the vicinity of 6--7~keV. 
In order to examine the existence of an emission line, 
we used only the data of obs.\ Q, which has much better statistics. 
We added a narrow Gaussian and fitted the spectrum, 
and then determined 
the center energy and the equivalent width of the Gaussian 
to be 
6.6 (6.4--6.9)~keV and 240 (40--440)~eV, respectively. 
However, the significance of the Gaussian is only $\sim 90$\% 
in the $F$-test, thus the existence of the emission line 
is still not clear. 
A confirmation of the existence of the emission line 
would further strengthen the X-ray binary nature of this source. 

\subsubsection{No.\ 105 --- AX~J0128.4$-$7329}

AX~J0128.4$-$7329 is located near 
the center of an expanding supergiant shell, 
SMC-1, which has a diameter of $\sim 1^\circ$ \citep{Meaburn1980}.
Wang and Wu (1992) reported that 
the X-ray emission from AX~J0128.4$-$7329 appears to be diffuse 
in an {Einstein} IPC image, and that 
the spectral hardness ratio indicates a temperature of 
$\sim 0.8$~keV if the absorption column density is assumed to be 
$3 \times 10^{20}$~cm$^{-2}$. 

Since the GIS spectrum has rather limited statistics, 
no rigid constraint was obtained for the spectral parameters. 
When we fitted the spectrum with a CIE thermal model, 
a temperature of 
2.6 (0.9--10) keV was obtained, which is marginally consistent with 
the argument of 
Wang and Wu (1992). 
Since half of the GIS FOV is polluted by stray light, which is 
probably from SMC~X-1, it is difficult to know 
the extent of AX~J0128.4$-$7329. 

\section{Discussion}

\subsection{Period Distribution of XBPs\label{sec:pdist}}
\begin{figure}
 \FigureFile(80mm,80mm){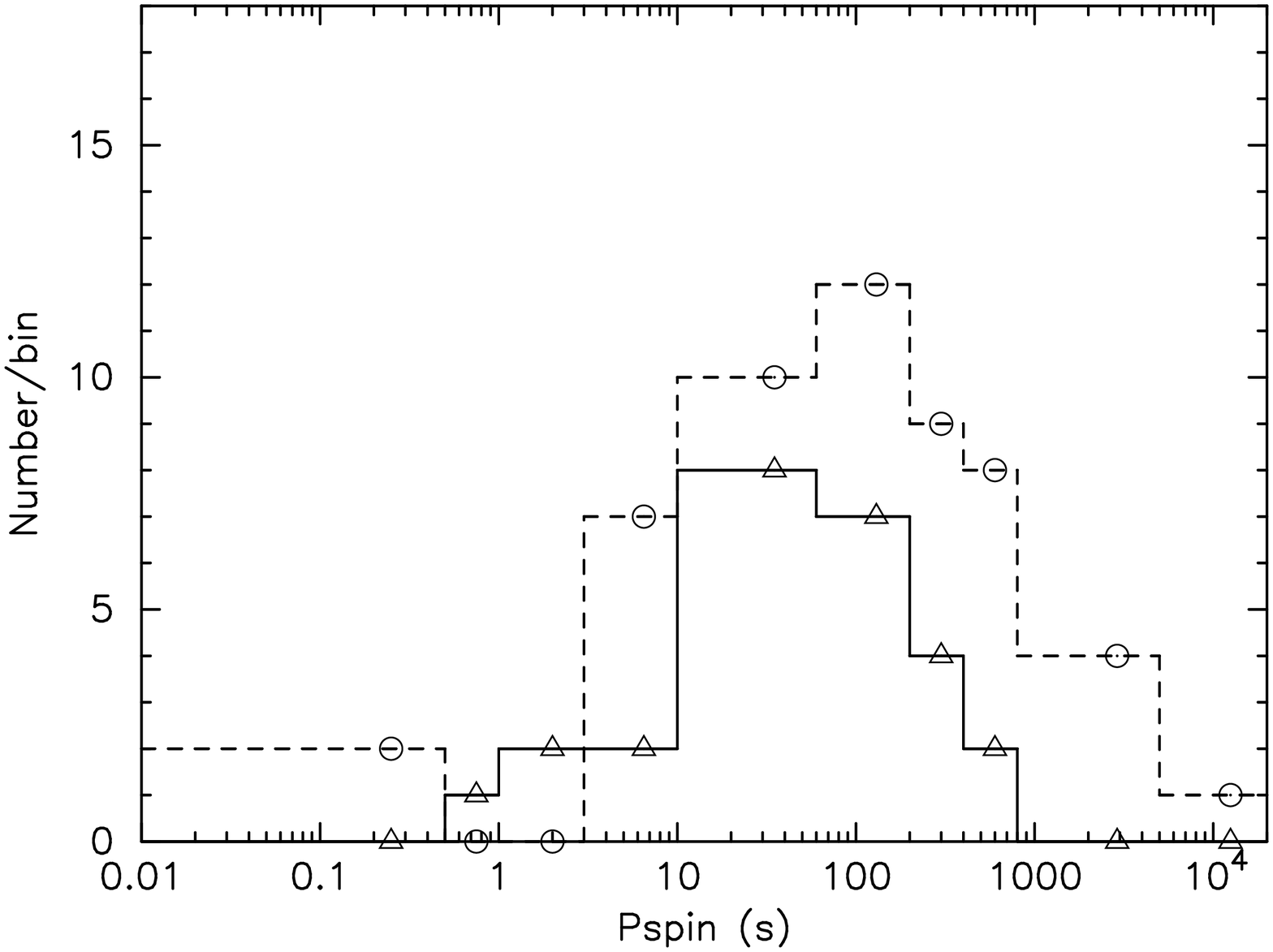}
 \caption{Period distribution of XBPs in the SMC 
(solid line; triangles) and 
our Galaxy (dashed line; circles). 
The horizontal and vertical axes, respectively, indicate 
the pulse period and the number of XBPs in each period bin. 
\label{fig:Pdist}}
\end{figure}

We have shown that 
at least 26 out of the 30 X-ray pulsars in the SMC are XBPs, 
probably with a high-mass star companion. 
Recently, \citet{Liu2000} 
compiled a catalogue of Galactic HMXBs 
in which 49 XBPs are included 
(in this paper we do not regard 4U~2206$+$543 as an XBP 
because of the non detection of pulsations in a further study 
by Corbet \& Peele, 2001). 
%
In addition, we investigated the recent literatures 
and found four new XBPs, 
AX~J1740.1$-$2847, 
XTE~J1543$-$568, 
SAX~J2239.3$+$6116, 
and AX~J1841.0$-$0536 
\citep{Sakano2000b,Finger2000,Intzand2001,Bamba2001}. 
In figure \ref{fig:Pdist} we show the distribution of pulse periods of 
the 53 XBPs in the Galaxy and 26 in the SMC.  
The average periods are $\sim 430$~s and $\sim 130$~s
for XBPs in the Galaxy and in the SMC, respectively. 
The significant difference is mainly caused by 
the lack of long period pulsars ($\gtrsim 1000$~s) in the SMC. 
%
%
%

We investigated the literature and found that 
Galactic XBPs with periods longer than 755~s (the longest period in the SMC) 
are all faint (table \ref{tab:longpulsars}). 
Their luminosities, $\lesssim 10^{35}$~erg~s$^{-1}$, 
correspond to $\lesssim 3 \times 10^{-13}$~erg~s$^{-1}$~cm$^{-2}$\ at the SMC distance, 
which are below, or only slightly above, the detection limit 
of this study. 
Therefore, we suggest that 
the lack of long-period pulsars in the SMC are 
merely due to the selection effect. 
This suggestion is supported by the fact that 
long-period pulsars in the SMC are also relatively faint 
(see subsubsections \ref{sec:755s} and \ref{sec:455s}). 
%
%

\subsection{Source Classification}
\subsubsection{Criteria for source classification\label{sec:class}}

We classify the 106 {ASCA} sources into several source classes, 
which are given in the ``Class'' column in table \ref{tab:cat_counter}. 
Definitions of the classes are the following. 

XBPs (``BP'' in table \ref{tab:cat_counter}) 
are the 18 pulsars having 
a long pulse period ($\sim 1$--1000~s), 
hard spectrum ($\Gamma \sim 1$), 
flux variability, and/or an optical counterpart, 
as described in subsection \ref{sec:pulsar}. 
%
No.\ 72 (CXOU~J0110043.1$-$721134) 
is proposed to be an AXP \citep{Lamb2002}; 
we thus designate this source as ``AXP''. 
Pulsars which are not definitely regarded as XBPs or AXPs 
are designated as ``P''. 
Thermal SNRs (``TS'') are the five SNRs from which 
emission lines of ionized atoms were detected in subsubsection \ref{sec:spec_snr}. 
The other SNRs with no significant emission line 
are classified as radio SNRs (``RS''). 
Candidates for XBPs and thermal SNRs (``BPc'' and ``TSc'') are 
defined in subsubsection \ref{sec:candidate}.

Nonpulsating HMXBs (``NH'') and candidates (``NHc'') are defined by 
the optical counterparts and flux variability 
as summarized in table \ref{tab:npHMXBs}. 
Grades from A to E are assigned according to 
the following criteria. 
Grade A and B sources are 
X-ray sources with a Be or a supergiant star companion. 
Flux variability has been known for grade A's, 
while not for grade B's. 
Grade C and D sources consist of 
X-ray sources having an emission-line object 
(Be star candidate) as a counterpart, 
which are catalogued in 
Haberl and Sasaki (2000). 
Again, flux variability has been known for grade C's 
while not for grade D's. 
Grade E sources also have an emission-line object counterpart, 
but other possibilities (AGN or active corona of a late 
type star) are proposed by 
Haberl and Sasaki (2000). 
%
In table \ref{tab:cat_counter}, 
we designate grade A and B sources as ``NH'' 
and grade C and D sources as ``NHc''.

Foreground stars (``FS'') and AGN (``AGN'') 
are defined merely based on positional coincidence of 
{ROSAT} sources of these classes 
(Sasaki et al.\ 2000 and references therein). 
Sources which do not fall into any classes 
are designated as ``UN'' (unclassified sources); 
UN(m) and UN(h) are defined in subsubsection \ref{sec:candidate}.

\subsubsection{Classification by hardness ratio\label{sec:classHR}}

In our earlier paper \citep{Yokogawa2000e}, 
we found a relation between 
source classes and their spectral hardness ratio (HR): 
%
XBPs and thermal SNRs fall in the regions of 
$0.2 \lesssim {\rm HR} \lesssim 0.6$ (``XBP region'') and 
$-1.0 \lesssim {\rm HR} \lesssim -0.6$ (``thermal SNR region''), 
respectively, 
while Crab-like SNRs and BH binaries  
stay between these regions (${\rm HR} \sim 0$). 
Having the increased number of sources, 
we investigated whether the same relation still holds. 
Figure \ref{fig:hr2} shows a plot of HR against the observed luminosity $L_{\rm obs}$, 
defined as $L_{\rm obs} = F_{\rm x} \times 4\pi d^2$,
where $d$ is the distance to the SMC (60~kpc). 
We find that seven XBPs and one thermal SNR that were newly added 
from the new data also satisfy the same relation. 
It is worth noting that 
most of the new XBPs and thermal SNR are 
relatively faint with $L_{\rm obs} \lesssim 10^{36}$~erg~s$^{-1}$, 
and that 
the classifications of these sources are mainly based on 
a very long {ASCA} observation or an {XMM-Newton} observation, 
which provides a better S/N ratio. 
This fact implies that the relation between source classes and HR 
is still valid for fainter sources.

\subsubsection{Candidates for XBPs, thermal SNRs, and background AGN\label{sec:candidate}}

\begin{figure}
 \FigureFile(160mm,80mm){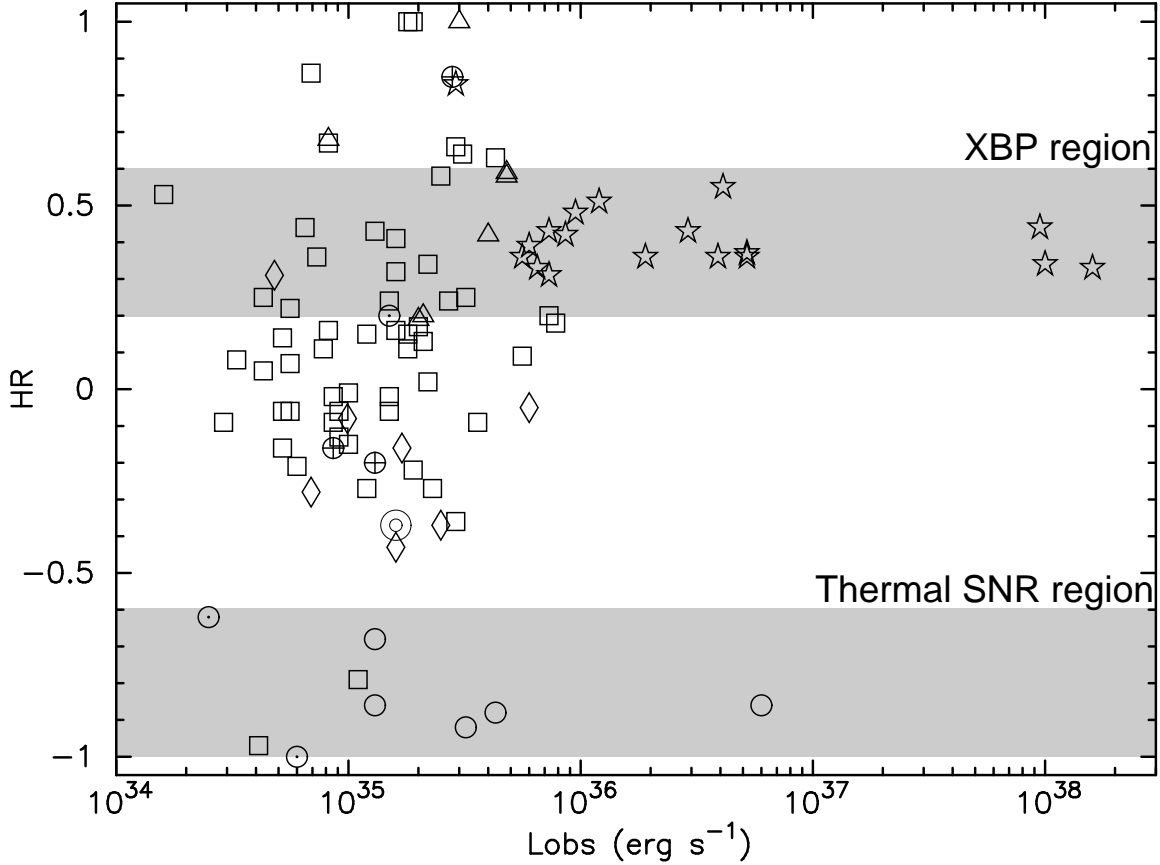}
 \caption{Plot of HR against $L_{\rm obs}$ (0.7--10.0~keV) 
for all classes of sources in the SMC. 
The symbols represent 
XBPs (stars), thermal SNRs (circles), 
nonpulsating HMXBs and their candidates (triangles), 
radio SNRs not regarded as thermal (circles with dots), 
pulsars not regarded as XBPs or AXPs (circles with a plus sign), 
AXP (double circle), 
sources coincident with AGN and foreground stars (diamonds), 
and other unclassified sources (squares). 
Sources detected multiple times are represented by 
single points, which correspond to 
the largest $L_{\rm obs}$ of each source. 
HR errors for the unclassified sources are mostly $\sim \pm 0.2$. 
%
Almost all XBPs and thermal SNRs fall in the 
``XBP region'' and ``thermal SNR region'', respectively. 

\label{fig:hr2}}
\end{figure}

Using the HR--$L_{\rm obs}$ plot (figure \ref{fig:hr2}), 
we regard sources 
in the XBP (or thermal SNR) region 
as candidates for XBPs (or thermal SNRs). 
We find 19 XBP candidates and four thermal SNR candidates, 
which are designated as BPc and TSc in table \ref{tab:cat_counter}, respectively. 
No.\ 91 is excluded from the XBP candidates 
because of the positional coincidence of an AGN. 
Although the HRs of No.\ 2 and No.\ 13%
\footnote{These sources are not plotted in figure \ref{fig:hr2} 
because $L_{\rm obs}$ is not derived (see subsubsection \ref{sec:spec_rem}).} 
are $-1$, we do not classify them as thermal SNRs 
because of their large HR errors. 
No.\ 25 also has a large HR error (${\rm HR} = -1.00 \pm 1.38$), 
but we classify this source as a thermal SNR 
because its SIS spectrum is typical of this class 
and because it has a radio SNR counterpart (0047$-$735).

Nonpulsating HMXBs and their candidates (table \ref{tab:npHMXBs}) 
are all located within or near the XBP region, 
thus are very promising targets for pulsation searches. 
Non-detection of pulsations from these sources is very likely 
due to the poor statistics. 
The detection of pulsations from low-flux sources 
AX~J0049.5$-$7323, AX~J0051.6$-$7311, 
and RX~J0101.3$-$7211 well demonstrates the necessity of 
high sensitivity observations 
(see subsubsections \ref{sec:755s}, \ref{sec:172s}, and \ref{sec:455s}). 

Two radio SNRs 
(0046$-$735 and 0047$-$735) 
are found among the thermal SNR candidates. 
The spectra of these sources have very limited statistics, 
and thus were not classified as thermal SNRs in subsubsection 
\ref{sec:spec_snr}. 
Another radio SNR, 0056$-$725, is located within 
the XBP region. Therefore, X-rays from this source may be attributed 
to an unresolved XBP in the radio SNR 0056$-$725, 
such as 1SAX~J0103.2$-$7209.

AGN and foreground stars 
are mostly located between the XBP and thermal SNR regions. 
Crab-like SNRs and BH binaries are also found at ${\rm HR} \sim 0$ 
\citep{Yokogawa2000e}. 
LMXBs have spectra of $\Gamma \sim 2$, 
which are similar to those of Crab-like SNRs; 
thus, LMXBs are also expected to be located at ${\rm HR} \sim 0$. 
We find 33 unclassified sources in this medium HR regime 
($-0.6 < {\rm HR} < 0.2$; ``UN(m)'' in table \ref{tab:cat_counter}; 
hereafter medium HR sources). 
Are they SMC members, or are they background/foreground sources?
To address this question, we used 
the $\log{N}$--$\log{S}$ relation from the {ASCA} Medium-Sensitivity Survey 
\citep{Ueda1999}. 
%
It is difficult to determine the accurate detection limit for our survey  
because of various uncertainties, such as 
contamination from bright sources, and the 
position-dependent efficiency of the detector. 
Therefore, we only make a rough estimation 
from figure \ref{fig:hr2} that the detection limit 
is $10^{-13}$--$10^{-12}$~erg~s$^{-1}$~cm$^{-2}$\ 
($4 \times 10^{34}$--$4 \times 10^{35}$~erg~s$^{-1}$). 
With this limit, several AGN could be detected in the GIS FOV 
in each observation; we can thus expect that 
several tens of AGN could be detected in our SMC survey. 
Therefore, it is likely that most of the 
33 medium HR sources are background AGN. 
This conclusion is independently supported by 
the spatial distribution of these sources, 
as shown in subsection \ref{sec:dist}.

The very hard regime (${\rm HR} > 0.6$) contains 
seven unclassified sources (``UN(h)'' in table \ref{tab:cat_counter}; 
hereafter, very hard sources). 
Even if a source has a hard spectrum with $\Gamma = 1.0$, 
a large absorption column of $N_{\rm H} > 10^{22}$~cm$^{-2}$\ is required 
for HR to exceed 0.6. 
Therefore, we regard these seven very hard sources as 
highly absorbed objects. 

\subsection{Source Populations in the SMC and in Our Galaxy}
\subsubsection{Basic data}
We have shown that 
most of the {ASCA} sources in the SMC region are 
classified as XBPs (mostly Be-XBPs), nonpulsating HMXBs, 
and SNRs, in addition to 
foreground stars and background AGN. 
In order to make a comparison of source populations 
in the SMC and in our Galaxy, 
we combined the results of our study with 
various catalogues compiled by other authors. 
The source classes included in this discussion are 
summarized in table \ref{tab:pop1}: 
HMXBs (XBPs and nonpulsating HMXBs), 
LMXBs, 
and 
SNRs (Crab-like and others).

HMXBs in the SMC consist of 26 XBPs (subsection \ref{sec:pulsar}) and 
eight nonpulsating HMXBs (grades A and B in table \ref{tab:npHMXBs}). 
Candidates for HMXBs are a combination of 
the 19 XBP candidates defined in this study (BPc in table \ref{tab:cat_counter}) 
and the grade C and D sources in table \ref{tab:npHMXBs}. 
Here, we regard the XBPs (and candidates) 
with an unknown optical counterpart as HMXBs, 
because most XBPs hitherto found are HMXBs 
(e.g., Bildsten et al.\ 1997). 
Evidence for LMXBs 
has not been detected from any source in the SMC. 
SNRs in the SMC have been surveyed by 
Mathewson et al.\ (1983, 1984) 
in the radio band 
and later by \citet{Filipovic1998a} in both the radio and X-ray bands; 
as a result 14 SNRs have been detected. 
Five SNR candidates have been found by \citet{Filipovic1998b}
and \citet{Haberl2000a}. 
In addition, two thermal SNR candidates defined in this study 
(No.\ 6 and No.\ 45) 
have no radio SNR counterpart, and are thus candidates for new SNRs. 
No evidence for Crab-like SNRs has been found.

HMXBs in our Galaxy are the 53 XBPs described in subsection \ref{sec:pdist} 
plus $\sim 30$ nonpulsating HMXBs catalogued in \citet{Liu2000}. 
Most complete LMXB catalogue is \citet{Liu2001}, 
in which $\sim 130$ LMXBs are contained. 
%
%
\citet{Green2000}
has compiled the most complete catalogue of 225 radio SNRs 
in our Galaxy. 
Among them, about 10 are 
associated with a rotation-powered X-ray pulsar, 
thus are regarded as Crab-like SNRs. 
In addition, some new SNRs have been discovered in the X-ray band 
with {ROSAT}, {ASCA}, {XMM-Newton}, and {Chandra}. 
We tentatively expect 100 SNR candidates to be discovered 
in the near future \citep{Aschenbach1996}.

\subsubsection{Comparison of the source populations\label{sec:comparison}}
Based on table \ref{tab:pop1}, we compare the source populations 
in the SMC and in our Galaxy from various aspects. 

Since the mass of the SMC is about 1/100th the mass of our Galaxy 
\citep{Westerlund1997}, 
the source numbers in our Galaxy should be divided by 100 
for a simple comparison 
(as in the second row of table \ref{tab:pop1}). 
As has been pointed out by several authors 
\citep{Schmidtke1999,Yokogawa2000e}, 
the normalized number of HMXBs 
is found to be much higher in the SMC. 
On the other hand, 
the normalized numbers of LMXBs are comparable in the two galaxies,
although the statistics are limited.  
Considering the detection limit of our survey, 
we conclude that 
the significant differences in the source numbers 
are not attributable to a selection effect 
(see subsection 5.4 of Yokogawa et al.\ 2000e for more detail). 
%
%
%
HMXBs are descendants of massive star binaries with ages of 
$\sim 10^7$~yr, while 
LMXBs comprise a much older population probably with ages of 
$\gtrsim 10^9$~yr. 
%
Therefore, we propose that 
the star forming rate was 
comparable between the two galaxies 
in a very old epoch ($\gtrsim 10^9$~yr ago), 
and then more recently ($\sim 10^7$~yr ago) 
it was much higher in the SMC.


%
HMXBs and type-II SNRs are both 
descendants of young massive stars, 
but the duration of being X-ray emitters would be highly different 
($\sim 10^6$--$10^7$~yr for the former 
while $\sim 10^5$~yr for the latter). 
Therefore, the number ratio between HMXBs and type-II SNRs 
probably indicates a change of the star formation rate 
in a very recent epoch ($\lesssim 10^7$~yr ago). 
We assume that SNRs in the SMC are all type-II 
because of the spatial distribution discussed in subsection \ref{sec:dist}, 
thus the number ratio of [HMXBs]/[type-II SNRs] 
(hereafter, H/II) 
in the SMC is $\approx 2$--3. 
%
%
As for our Galaxy, 
H/II is estimated to be $\approx 0.5$--0.7, 
if we simply assume that 50\% of Galactic SNRs are type-II. 
This assumption is reasonable because 
many SNRs are concentrated in the Galactic plane 
(e.g., $\sim 60$\% are at $|b| < 1^\circ$; Green 2000), 
and are thus considered to be mainly type-II. 
The much larger value of H/II in the SMC implies 
that there was a dramatic decline of the star-forming rate
in a very recent epoch ($\lesssim 10^7$~yr ago), 
following active star formation 
of $\sim 10^7$~yr ago described above. 

Although 
both HMXBs and Crab-like SNRs are direct descendants of massive stars, 
the number ratios of these classes are significantly different: 
10/83 in our Galaxy 
and 0/34 in the SMC (0/74 when the HMXB candidates are included). 
If the ratio is identical between galaxies, 
at least several Crab-like SNRs should be found in the SMC. 
%
Crab-like SNRs in the SMC, if exist, would be found 
in the medium HR regime (${\rm HR} \sim 0$), 
where those in the LMC are located \citep{Yokogawa2000e}. 
Therefore, some of the 33 medium HR sources may be Crab-like SNRs; 
non-detection of pulsations could be due to limited statistics. 
High-sensitivity observations 
of those sources with a good time resolution would thus be fruitful. 
In addition, high-resolution radio surveys 
to search for plerionic emission are also encouraged. 
If, on the other hand, Crab-like SNRs are really lacking in the SMC, 
it may imply that 
the binary frequency of massive stars is much higher in the SMC 
because Crab-like SNRs and HMXBs are 
descendants of single and binary massive stars, respectively. 
A higher binary frequency can in part be 
a cause of the higher number ratio of HMXBs to type-II SNRs 
mentioned above.


\subsection{Spatial Distribution of Various Classes of Sources\label{sec:dist}}

We investigated the spatial distributions of 
four classes of sources: 
HMXBs, SNRs, AGN plus medium HR sources, and very hard sources. 
In this subsection, ``HMXBs'' include 
the XBPs, nonpulsating HMXBs, and HMXB candidates in table \ref{tab:pop1},
while ``SNRs'' include 
the Crab-like and other SNRs and SNR candidates in table \ref{tab:pop1}.  
Medium HR sources and very hard sources are defined in 
subsubsection \ref{sec:candidate}.

\begin{figure}
 \FigureFile(80mm,80mm){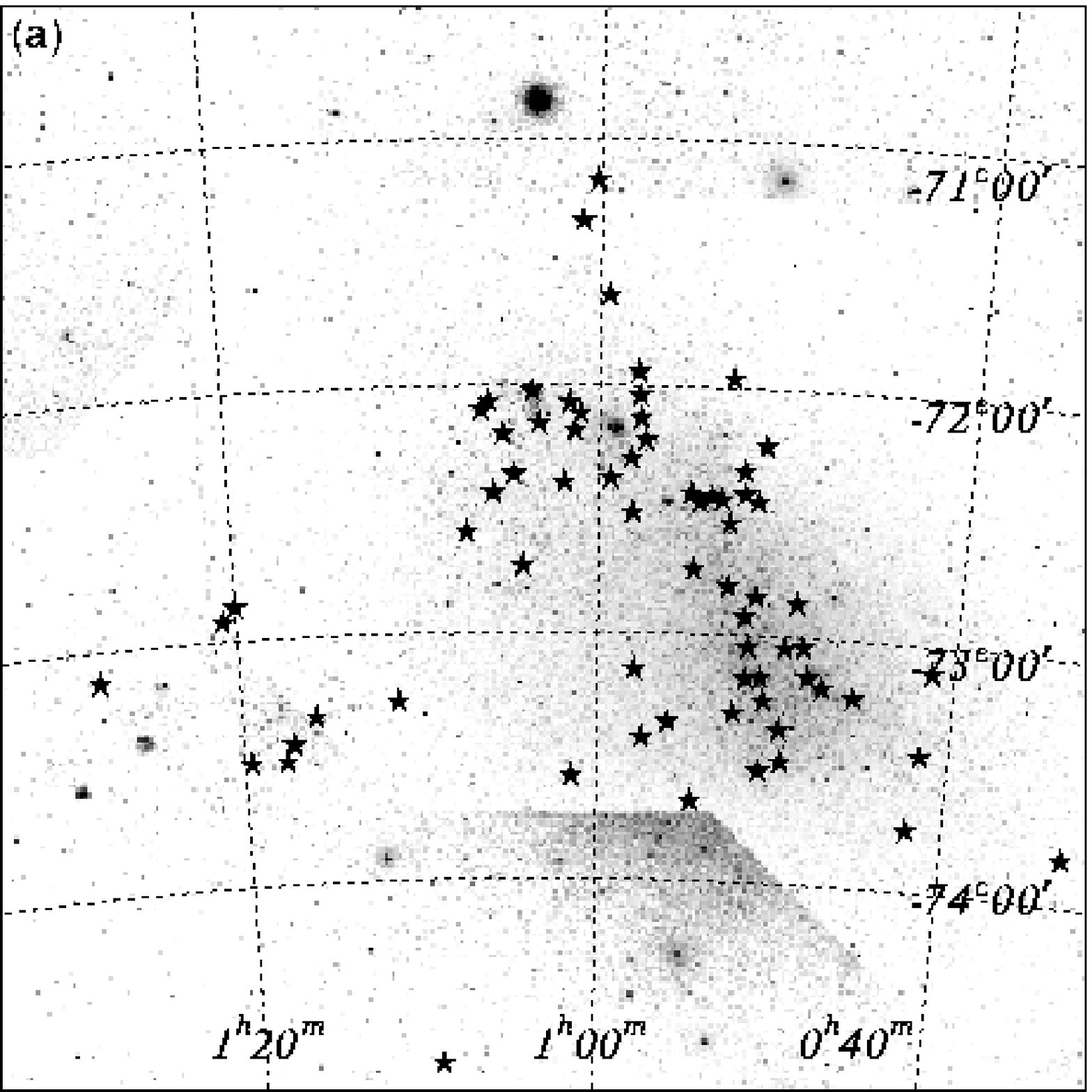}
 \FigureFile(80mm,80mm){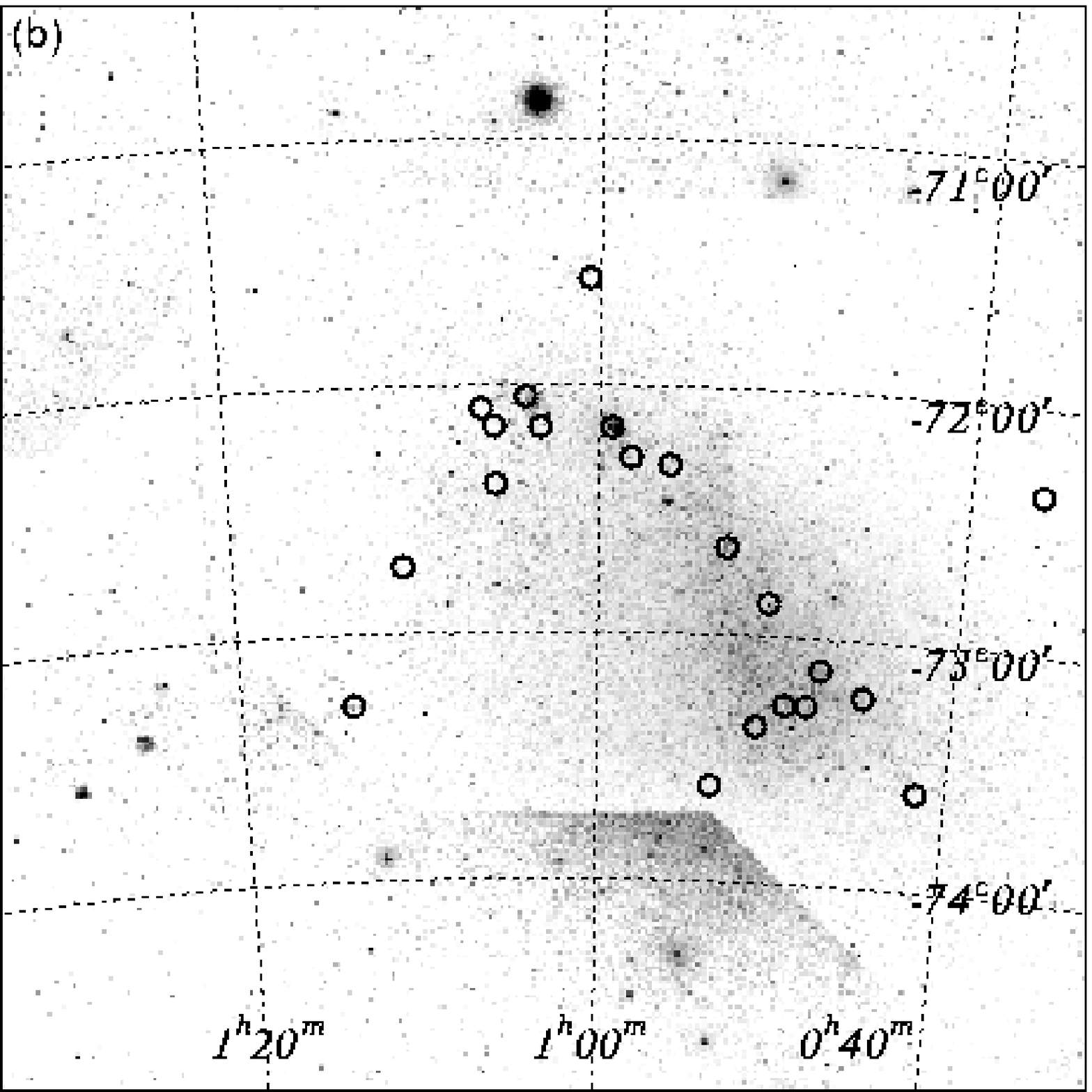}

 \FigureFile(80mm,80mm){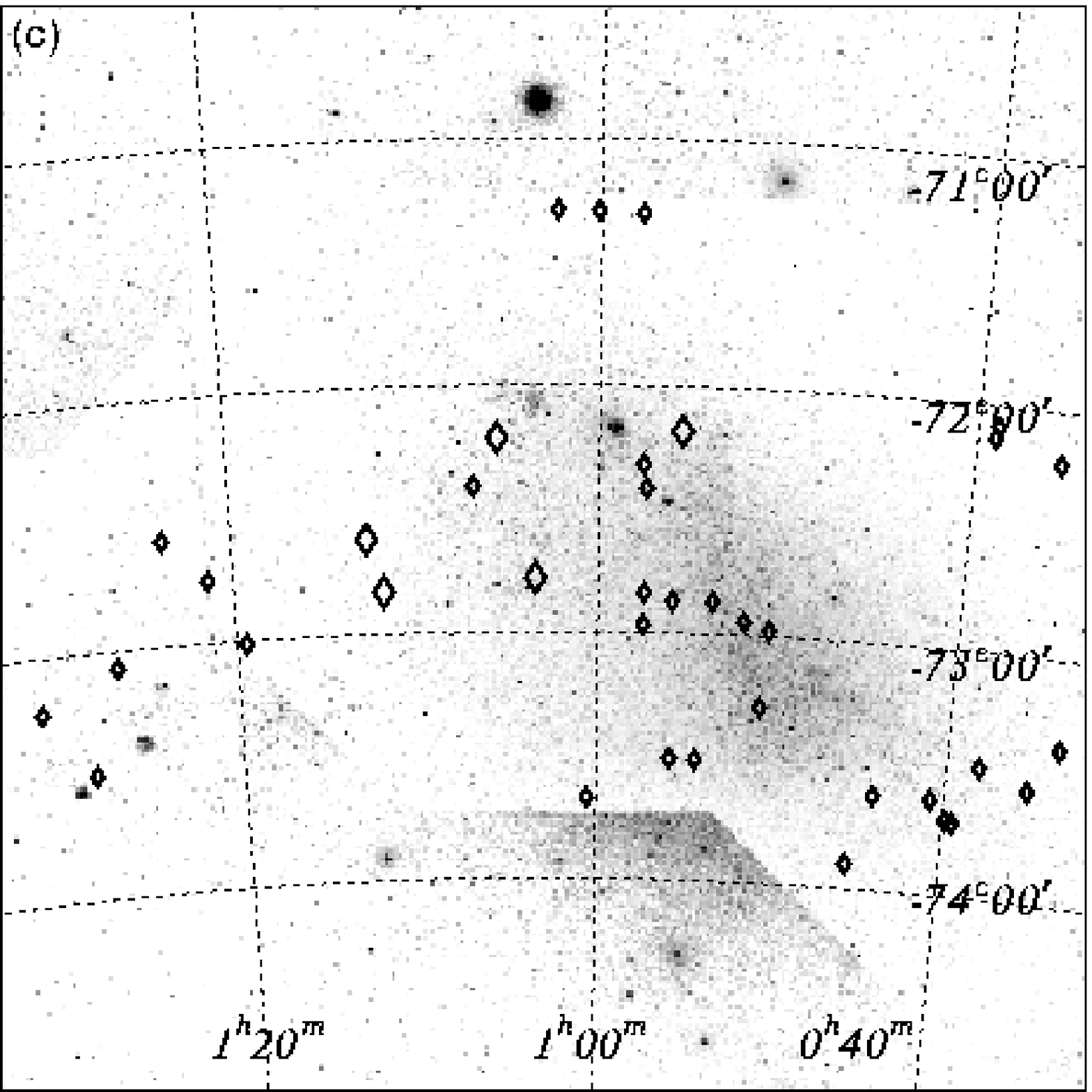}
 \FigureFile(80mm,80mm){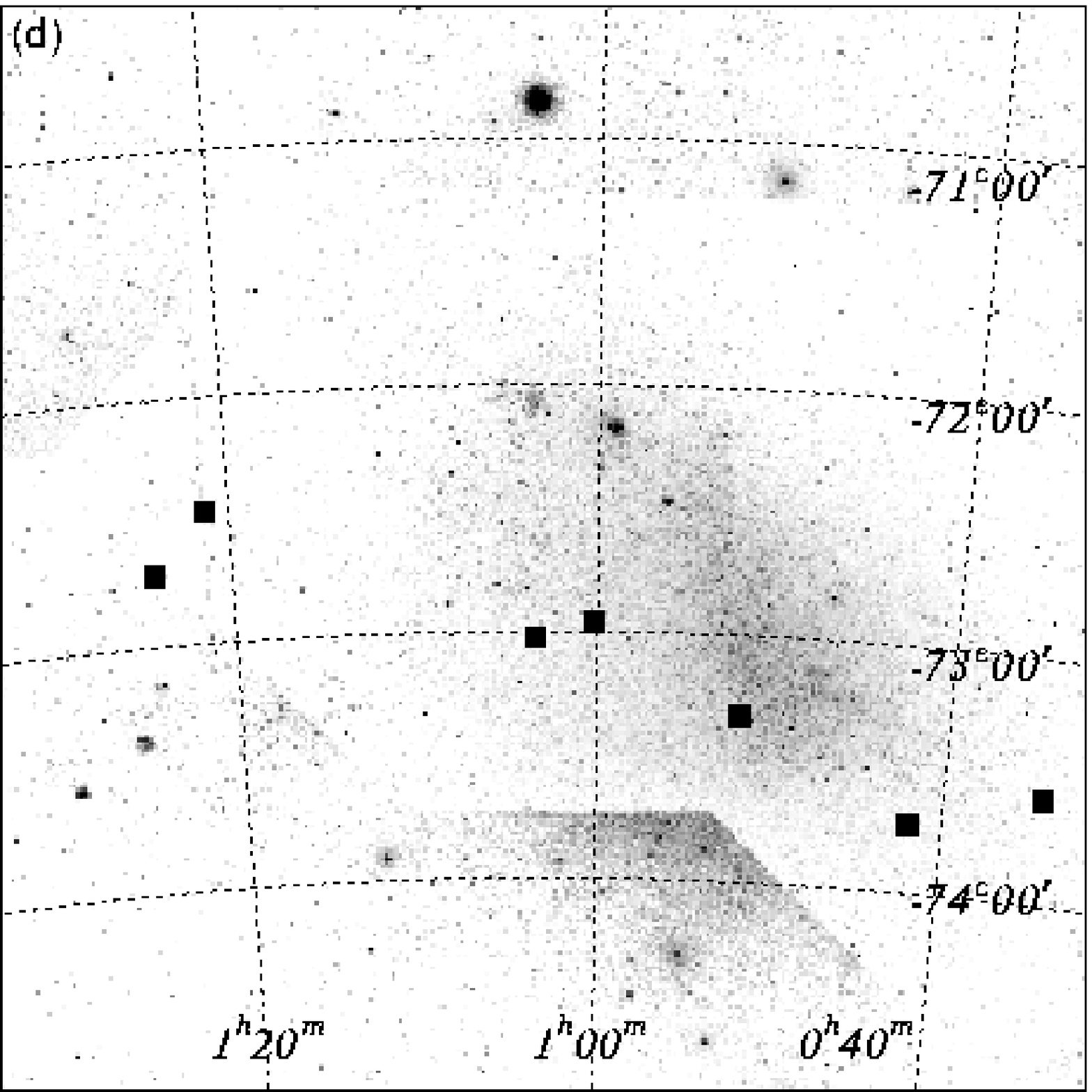}
 \caption{Spatial distribution of HMXBs 
(a: indicated by stars), 
SNRs (b: circles), 
and 
the six AGN candidates and the 33 medium HR sources 
(c: diamonds), 
and the seven very hard sources (d: squares), 
superimposed on the DSS image of the SMC.
The large and small diamonds in (c) indicate 
the AGN and the medium HR sources, respectively. 
The discontinuity seen in the DSS image is an artifact. 
The equatorial coordinates in an equinox 2000 are overlaid.
\label{fig:dist}}

\end{figure}
\begin{figure}
 \FigureFile(80mm,80mm){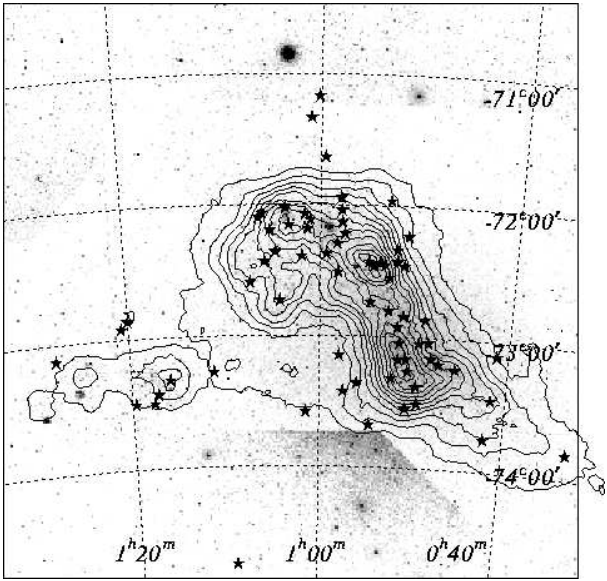}
 \caption{Spatial distribution of HMXBs (figure \ref{fig:dist}a) 
superimposed on the isodensity map of 
young stars with ages of (1.2--$3) \times 10^7$~yr 
(contours; taken from 
Maragoudaki et al.\ 2001). 
The discontinuity seen in the DSS image is an artifact. 
The equatorial coordinates in an equinox 2000 are overlaid.
\label{fig:HMXB_smcmorrev5}}
\end{figure}

We show the spatial distribution of HMXBs in figure \ref{fig:dist}a. 
We find that most HMXBs are concentrated in the optical main body, 
and $\sim 10$\% are located in the eastern wing. 
\citet{Maragoudaki2001} carried out an optical survey of the SMC 
and derived spatial distributions of stars 
in seven ranges of ages 
from $> 2\times 10^9$~yr to $< 8 \times 10^6$~yr. 
They found that very old stars have a smooth and spheroidal distribution, 
while younger stars are concentrated in the main body and the eastern wing. 
Comparing figure \ref{fig:dist}a with their results, 
we find that the distribution of the HMXBs 
well resembles that of younger stars, 
especially stars with ages of (1.2--$3) \times 10^7$~yr 
(figure \ref{fig:HMXB_smcmorrev5}). 
This is naturally expected 
because HMXBs are descendants of massive (young) stars. 

The spatial distribution of SNRs shown in figure \ref{fig:dist}b 
is very similar to that of HMXBs. 
Therefore, we regard that 
most of the SNRs are descendants of massive stars (i.e., type-II SNRs). 
This is consistent with 
the fact that so far type-Ia SNRs have not been found in the SMC, 
while three SNRs have type-II origin (subsection \ref{sec:snr}).

We also investigated the distribution of the 
33 medium HR sources 
and five AGN 
as shown in figure \ref{fig:dist}c. 
The distribution is relatively uniform, 
which exhibits a clear contrast with HMXBs and SNRs 
and is qualitatively consistent with the distribution of 
old stars ($> 2 \times 10^9$~yr) 
found by \citet{Maragoudaki2001}. 
%
This fact implies 
either that
(1) most of the medium HR sources are unrelated to the SMC, i.e., 
background or foreground sources, 
or that 
(2) they represent a population much older than the HMXBs. 
Scenario (1) is consistent with the estimate 
from the $\log{N}$--$\log{S}$ relation 
discussed in subsubsection \ref{sec:candidate}. 
However, taking account of the large uncertainty in the above estimate, 
there still remains a possibility 
that some of the medium HR sources represent 
an older population, probably LMXBs. 
Since the number of LMXBs has a relatively large impact 
on the star-forming activity in the old epoch 
(see subsubsection \ref{sec:comparison}), 
follow-up observations of these medium HR sources are encouraged.

The distribution of the seven very hard sources 
shown in figure \ref{fig:dist}d 
seems to be as uniform as 
that of the AGN and the medium HR sources, 
although the number of sources is highly limited. 
Therefore, the very hard sources may be 
unrelated to the SMC, i.e., 
background or foreground sources. 
Considering the suggestion that these sources should be 
highly absorbed (see \ref{sec:candidate}), 
we propose that most of them are Seyfert 2 galaxies. 
We thus encourage optical follow-up observations of these sources.

J.Y., K.I., and M.T.\ were financially
supported by JSPS Research Fellowship for Young Scientists.
We are grateful to Prof.\ Fukazawa 
for his help when revising the manuscript. 
We retrieved ROSAT data from the HEASARC Online System
which is provided by NASA/GSFC.

\onecolumn

%

\clearpage

\clearpage

\begin{table}
\caption{{ASCA} observations of the SMC region. \label{tab:obs}}
\begin{tabular}{clllccc}
\hline\hline
\multicolumn{1}{c}{Obs.\ ID} & 
\multicolumn{2}{c}{Observation date (UT)} &
\multicolumn{1}{c}{Target name} &
\multicolumn{2}{c}{Pointed direction (J2000)} & 
\multicolumn{1}{c}{Exposure (ks)} \\
 &\multicolumn{1}{c}{Start} &\multicolumn{1}{c}{End} & & 
R.A. ({$^{\rm h}$}\ $^{\rm m}$\ $^{\rm s}$)  & Dec. ($^\circ$ $'$ $''$) & GIS/SIS \\
\hline
A&1993/04/26 22:23&04/27 23:41&           SMC X-1&1 17 04&$-$73 26 34&  4/ --- \\
B&1993/05/12 08:01&05/13 12:26&      E0102$-$72.2&1 04 02&$-$72 01 55& 35/31\rlap{$^{a}$}\\
C&1995/10/18 18:32&10/19 21:31&           SMC X-1&1 17 04&$-$73 26 35& 41/--- \\
D&1996/05/21 02:25&05/23 06:31&       0103$-$72.6&1 04 52&$-$72 23 11& 78/61\rlap{$^{a}$}\\
E&1997/05/21 13:01&05/22 01:31&  1E 0035.4$-$7230&0 37 20&$-$72 14 14& 23/--- \\
F&1997/11/13 03:58&11/14 06:51&               N19&0 47 16&$-$73 08 30& 43/38\rlap{$^{b}$}\\
\end{tabular}
\end{table}

\begin{table}
\caption{{ASCA} counterparts for {ROSAT} sources selected with 
the conservative criteria. \label{tab:counter_abs}}
\begin{tabular}{rcccrrl}
\hline\hline
\multicolumn{2}{c}{--- {ROSAT}\rlap{$^{a}$} ---}& 
\multicolumn{3}{c}{------------ {ASCA}\rlap{$^{b}$} ------------} & & \\
\multicolumn{1}{c}{No.} & 
\multicolumn{1}{c}{Error} & 
\multicolumn{1}{c}{No.} & 
\multicolumn{1}{c}{Obs.\ ID} & 
\multicolumn{1}{c}{Off-axis}& 
\multicolumn{1}{c}{Sep\rlap{$^{c}$}} & 
\multicolumn{1}{c}{Remarks} \\
&
\multicolumn{1}{c}{($''$)} &
& &
\multicolumn{1}{c}{($'$)} &
\multicolumn{1}{c}{($''$)} & \\
\hline
  34 & 1.8 & 32 & F & 17.0 & 31.9 & Be-XBP AX~J0051$-$733 ($P=323$~s)\\
  34 & 1.8 & 32 & Q &  9.7 & 24.0 & Be-XBP AX~J0051$-$733 ($P=321$~s)\\
  37 & 2.6 & 37 & H & 17.2 & 45.5 & Be-XBP AX~J0051$-$722 ($P=91$~s)\\
  38 & 2.7 & 36 & Q & 14.3 & 40.2 & SNR 0049$-$736\\
  41 & 1.5 & 40 & Q & 11.7 & 32.6 & Be-XBP AX~J0051.6$-$7311 ($P=172$~s)\\
\end{tabular}
\end{table}

\clearpage

\renewcommand{\arraystretch}{0.8}
\begin{table}
 \caption{{ASCA} sources detected in multiple observations.
\label{tab:asca-asca}}
%
}
\end{landscape}

\clearpage

\begin{table}
\setlength{\tabcolsep}{10pt} 
\caption{Luminosities of the long period pulsars in our Galaxy. 
\label{tab:longpulsars}}
\begin{tabular}{lrcrc}
\hline\hline
\multicolumn{1}{c}{Name}&
\multicolumn{1}{c}{Period}&
\multicolumn{1}{c}{Distance}&
\multicolumn{1}{c}{{$L_{\rm X}$}}&
\multicolumn{1}{c}{References}\\
&
\multicolumn{1}{c}{(s)}&
\multicolumn{1}{c}{(kpc)}&
\multicolumn{1}{c}{(erg~s$^{-1}$)}& 
\\
\hline
X Per 			&835\phantom{.2}	&0.95$^{1}$	&$<(2-10) \times 10^{33}$&1\\
RX J1037.5$-$5647	&862\phantom{.2}	&5$^{2}$\phantom{.95}	&$1 \times 10^{33}$	&2\\
SAX J2239.3$+$616	&1247.2		&4.4$^{3}$\phantom{5}	&$9 \times 10^{32}$	&3\\
RX J0146.9$+$6121	&1404.2		&2.5$^{4}$\phantom{5}	&$<(2-5) \times 10^{33}$&4\\
2S 0114+650		&10008\phantom{.2}	&7.2$^{5}$\phantom{5}	&$6 \times 10^{34}$	&5\\
\hline
\multicolumn{5}{p{0.98\textwidth}}{
Note. --- Luminosities are converted from 
fluxes presented by \citet{Liu2000}, assuming the distances 
in each reference.
}\\
\multicolumn{5}{p{0.98\textwidth}}{
References --- 
(1) \citet{Telting1998}; 
(2) \citet{Motch1997}; 
(3) \citet{Intzand2000}; 
(4) \citet{Haberl1998}; 
(5) \citet{Reig1996}
}
\end{tabular}
\end{table}
%

\clearpage

\scriptsize
\centering
\setlength{\tabcolsep}{6pt}
\renewcommand{\arraystretch}{1.6}
\begin{longtable}{ccclccc}
 \caption{Nonpulsating HMXBs and likely candidates in the SMC. 
\label{tab:npHMXBs}}
\hline\hline
\multicolumn{2}{c}{--- Number\rlap{$^{a}$} ---}&
\multicolumn{1}{c}{Name}&
\multicolumn{1}{c}{Optical\rlap{$^{b}$}}&
\multicolumn{1}{c}{Variable?\rlap{$^{c}$}}&
\multicolumn{1}{c}{Grade\rlap{$^{d}$}}&
\multicolumn{1}{c}{References} \\
\multicolumn{1}{c}{{ASCA}} &
\multicolumn{1}{c}{HRI} & 
 &
\multicolumn{1}{c}{identification} &
 &
 & \\
\hline
\endhead 
\hline
\hline
\multicolumn{7}{l}{\hbox to 0pt{\parbox{150mm}{\footnotesize 
a: Source number in the catalogues of 
{ASCA} (table \ref{tab:cat_counter}) and HRI \citep{Haberl2000b}. }}} \\
\multicolumn{7}{l}{\hbox to 0pt{\parbox{150mm}{\footnotesize 
b: Spectral type of the optical counterpart. ``Be?'' indicates 
an emission line object catalogued in 
Meyssonnier and Azzopardi (1993)
or 
Murphy and Bessell (2000).  
``SG'' indicates a supergiant counterpart.  }}} \\
\multicolumn{7}{l}{\hbox to 0pt{\parbox{150mm}{\footnotesize 
c: For sources with the ``Yes'' sign, 
long-term X-ray variability has been found.  }}} \\
\multicolumn{7}{l}{\hbox to 0pt{\parbox{150mm}{\footnotesize 
d: In table \ref{tab:cat_counter}, we designate grade A and B sources 
as NH and grade C and D sources as NHc. 
Grade E sources are not regarded as non-pulsating HMXBs 
(see subsubsection \ref{sec:class}). }}} \\
\multicolumn{7}{l}{\hbox to 0pt{\parbox{150mm}{\footnotesize 
e: This source is designated as RX~J0104.5$-$7121  
in 
Haberl and Sasaki (2000) 
(subsection 6.4), 
but probably should be RX~J0104.5$-$7\underline{2}21 
according to its coordinate.  }}} \\
\multicolumn{7}{l}{\hbox to 0pt{\parbox{150mm}{\footnotesize 
f: A point-like radio source is also located in the error 
circle, thus the possibility that it is an AGN remains \citep{Haberl2000b}. }}} \\
\multicolumn{7}{l}{\hbox to 0pt{\parbox{150mm}{\footnotesize 
g: X-ray emission may be attributable 
to an active corona (AC) of a late type star \citep{Haberl2000b}.  }}} \\
\multicolumn{7}{l}{\hbox to 0pt{\parbox{150mm}{\footnotesize 
References --- 
(1) Haberl and Sasaki (2000); 
(2) Kahabka and Pietsch (1996);
(3) \citet{Crampton1978};
(4) \citet{Clark1978};
(5) Whitlock and Lochner (1994);
(6) Wang and Wu (1992);
(7) this work;
(8) \citet{Filipovic2000b};
 }}}\\
\endlastfoot
 ...&  1&RX~J0032.9$-$7348	&Be\rlap{$^{1}$}    &Yes\rlap{$^{2}$}     &A &1,2\\ 
 ...& 22&SMC~X-3		&Be\rlap{$^{3}$}    &Yes\rlap{$^{4}$}     &A &3,4\\ 
 ...&...&1H 0103$-$762		&Be\rlap{$^{5}$}    &Yes\rlap{$^{2}$}     &A &2,5\\ 
 ...& 42&RX~J0101.0$-$7206	&Be\rlap{$^{1}$}    &Yes\rlap{$^{2}$}     &A &1,2\\ 
 ...&...&EXO~0114.6$-$7361	&B (SG)\rlap{$^{6}$}&Yes\rlap{$^{6}$}     &A &6\\ 
 ...& 24&2E~0051.1$-$7304	&Be\rlap{$^{1}$}    &Yes?\rlap{$^{6}$}        &B &1,6\\ 
  63& 38&RX~J0058.2$-$7231	&Be\rlap{$^{1}$}    &...           &B &1\\ 
 ...&...&RX~J0106.2$-$7205	&Be\rlap{$^{1}$}    &...           &B &1\\ 
%
  22&...&AX~J0048.2$-$7309      &Be?\rlap{$^{8}$}   &Yes\rlap{$^{8}$}     &C &7\\ 
  50& 30&RX~J0054.9$-$7245	&Be?\rlap{$^{1}$}   &Yes\rlap{$^{1}$}     &C &1\\ 
 ...& 36&RX~J0057.8$-$7207	&Be?\rlap{$^{1}$}   &Yes\rlap{$^{1}$}     &C &1\\ 
 ...& 44&RX~J0101.6$-$7204	&Be?\rlap{$^{1}$}   &Yes\rlap{$^{1}$}     &C &1\\ 
  76& 45&RX~J0101.8$-$7223	&Be?\rlap{$^{1}$}   &Yes\rlap{$^{1}$}     &C &1\\ 
 ...& 50&RX~J0103.6$-$7201	&Be?\rlap{$^{1}$}   &Yes\rlap{$^{1}$}     &C &1\\ 
 ...& 53&RX~J0105.1$-$7211	&Be?\rlap{$^{1}$}   &Yes\rlap{$^{9}$}     &C &1,8\\ 
  87& 56&RX~J0107.1$-$7235	&Be?\rlap{$^{1}$}   &Yes\rlap{$^{1}$}     &C &1\\ 
 ...&  2&RX~J0041.2$-$7306	&Be?\rlap{$^{1}$}   &...           &D &1\\ 
 ...&  3&RX~J0045.6$-$7313	&Be?\rlap{$^{1}$}   &...           &D &1\\ 
 ...&  8&RX~J0048.5$-$7302	&Be?\rlap{$^{1}$}   &...           &D &1\\ 
  28& 11&RX~J0049.5$-$7331	&Be?\rlap{$^{1}$}   &Yes?\rlap{$^{8}$}        &D &1,7\\ 
 ...& 14&RX~J0050.7$-$7332	&Be?\rlap{$^{1}$}   &...           &D &1\\ 
  35& 15&RX~J0050.9$-$7310	&Be?\rlap{$^{1}$}   &...           &D &1\\ 
 ...& 17&RX~J0051.3$-$7250	&Be?\rlap{$^{1}$}   &...           &D &1\\ 
 ...& 21&No.\ 26 in WW92	&Be?\rlap{$^{1}$}   &...           &D &1\\ 
 ...& 26&RX~J0053.4$-$7227	&Be?\rlap{$^{1}$}   &...           &D &1\\ 
 ...& 29&RX~J0054.5$-$7228	&Be?\rlap{$^{1}$}   &...           &D &1\\ 
 ...& 32&2E~0054.4$-$7237 (IKT13)&Be?\rlap{$^{1}$}  &...           &D &1\\ 
 ...& 34&2E~0055.8$-$7229 (IKT15)&Be?\rlap{$^{1}$}  &...           &D &1\\ 
 ...& 37&RX~J0057.9$-$7156	&Be?\rlap{$^{1}$}   &...           &D &1\\ 
 ...& 51&RX~J0104.1$-$7243	&Be?\rlap{$^{1}$}   &...           &D &1\\ 
 ...& 52&RX~J0104.5$-$7121\rlap{$^{e}$}&Be?\rlap{$^{1}$}   &...           &D &1\\ 
 ...& 54&RX~J0105.7$-$7226	&Be?\rlap{$^{1}$}   &...           &D &1\\ 
  85& 55&RX~J0105.9$-$7203	&Be?\rlap{$^{1}$}   &...           &D &1\\ 
 ...& 60&RX~J0119.6$-$7330	&Be?\rlap{$^{1}$}   &...           &D &1\\ 
 ...&  5&RX~J0047.3$-$7239	&Be?AGN?\rlap{$^{f,1}$} &...       &E &1\\ 
 ...& 18&RX~J0051.8$-$7159&Be?AC?\rlap{$^{g,1}$}    &...           &E &1\\ 
\end{longtable}

\clearpage

\begin{table}
 \caption{Source populations in the SMC and in our Galaxy. 
\label{tab:pop1}}
\begin{tabular}{lccccccc}
\hline\hline
&\multicolumn{3}{c}{------------ HMXBs ------------}
&\multicolumn{1}{c}{LMXBs}
&\multicolumn{3}{c}{------------ SNRs ------------}\\
&
\multicolumn{1}{c}{XBPs}&
\multicolumn{1}{c}{Nonpulse}&
\multicolumn{1}{c}{Candidates}&
&
\multicolumn{1}{c}{Crab-like}&
\multicolumn{1}{c}{Others}&
\multicolumn{1}{c}{Candidates}\\
\hline 
Galaxy      & 53   & 30   & ...        & 130   & 10  & 220   & 100 \\  
Galaxy $\times 1/100$\rlap{$^{a}$}
        &  0.53&  0.3 & ...        &   1.3 &  0.1&   2.2 &   1 \\  
SMC         & 26   &  8   & 40         &   0   &  0  &  14   &   7  \\
\hline 
\multicolumn{8}{p{0.9\textwidth}}{a: Since the mass of the SMC is about 1/100 of 
that of our Galaxy, the source numbers in our Galaxy should be divided 
by 100 for a simple comparison.}\\
\end{tabular}
\end{table}


\begin{thebibliography}{}
\bibitem[Aschenbach(1996)]{Aschenbach1996} 
 Aschenbach, B. 1996, MPE Report, 263, 213
\bibitem[Bamba et al.(2001)]{Bamba2001}
 Bamba, A., Yokogawa, J., Ueno, M., Koyama, K., \& Yamauchi, S. 
 2001, \pasj, 53, 1179
\bibitem[Bildsten et al.(1997)]{Bildsten1997}
 Bildsten, L., Balser, D. S., Chiu, J., Finger, M. H., 
 Koh, D. T., Nelson, R. W., Prince, T. A., Rubin, B. C., et al. 1997, 
 \apjs, 113, 367
\bibitem[Buckley et al.(2001)]{Buckley2001}
 Buckley, D. A. H., Coe, M. J., Stevens, J. B., 
 van der Heyden, K., Angelini, L., White, N., \& Giommi, P. 
 2001, \mnras, 320, 281
\bibitem{Burke1994} 
 Burke, B. E., Mountain, R. W., Daniels, P. J., Cooper, M. J., \& Dolat, V. S. 
 1994, IEEE Trans. Nucl. Sci., 41, 375
\bibitem[Chakrabarty et al.(1998a)]{Chakrabarty1998a} 
 Chakrabarty, D., Levine, A. M., Clark, G. W. \& Takeshima, T. 
 1998a, \iaucirc, 7048
\bibitem[Chakrabarty et al.(1998b)]{Chakrabarty1998b} 
 Chakrabarty, D., Takeshima, T., Ozaki, M., Paul, B., \& Yokogawa, J. 
 1998b, \iaucirc, 7062
\bibitem[Clark et al.(1978)]{Clark1978} 
 Clark, G., Doxsey, R., Li, F., Jernigan, J. G., \& 
 van Paradijs, J. 1978, \apj, 221, L37
\bibitem[Clark et al.(1996)]{Clark1996}
 Clark, G., Remillard, R., \& Woo, J. 1996, \iaucirc, 6282
\bibitem[Clark et al.(1997)]{Clark1997}
 Clark, G. W., Remillard, R. A., \& Woo, J. W. 1997, \apj, 474, L111 
\bibitem[Coe et al.(1998)]{Coe1998}
 Coe, M. J., Buckley, D. A. H., Charles, P. A., 
 Southwell, K. A., \& Stevens, J. B. 1998, \mnras, 293, 43
\bibitem[Coe et al.(2000)]{Coe2000a}
 Coe, M. J., Haigh, N. J., \& Reig, P. 
 2000, \mnras, 314, 290
\bibitem[Coe et al.(2002)]{Coe2002}
 Coe, M. J., Haigh, N. J., Laycock, S. G. T., 
 Negueruela, I., \& Kaiser, C. R. 2002, \mnras, 332, 473
\bibitem[Coe, Orosz(2000)]{Coe2000b}
 Coe, M. J., \& Orosz, J. A. 2000, \mnras, 311, 169
\bibitem[Corbet(1984)]{Corbet1984}
 Corbet, R. H. D. 1984, \aap, 141, 91
\bibitem[Corbet et al.(2002)]{Corbet2002}
 Corbet, R., Markwardt, C. B., Marshall, F. E., Laycock, S., 
 \& Coe, M. 2002, \iaucirc, 7932
\bibitem[Corbet et al.(2001b)]{Corbet2001b}
 Corbet, R. H. D., Marshall, F. E., Coe, M. J., Laycock, S., 
 \& Handler, G. 2001b, \apj, 548, L41
\bibitem[Corbet et al.(1998)]{Corbet1998} 
 Corbet, R., Marshall, F. E., Lochner, J. C., 
 Ozaki, M., \& Ueda, Y. 1998, \iaucirc, 6803
\bibitem[Corbet et al.(2001a)]{Corbet2001a}
 Corbet, R., Marshall, F. E., \& Markwardt, C. B.
 2001a, \iaucirc, 7562
\bibitem[Corbet, Peele(2001)]{Corbet2001c}
 Corbet, R. H. D., \& Peele, A. G. 2001, \apj, 562, 936
\bibitem[Cowley et al.(1997)]{Cowley1997} 
 Cowley, A. P., Schmidtke, P. C., McGrath, T. K., Ponder, A. L., 
 Fertig, M. R., Hutchings, J. B., \& Crampton, D. 
 1997, \pasp, 109, 21
\bibitem[Crampton et al.(1978)]{Crampton1978}
 Crampton, D., Hutchings, J. B., \& Cowley, A. P. 
 1978, \apj, 223, L79
\bibitem[Cusumano et al.(2000)]{Cusumano2000}
 Cusumano, G., Maccarone, M. C., Nicastro, L., Sacco, B., 
 \& Kaaret, P. 2000, \apj, 528, L25
\bibitem[Davies et al.(1976)]{Davies1976}
 Davies, R. D., Elliott, K. H., \& Meaburn, J.\ 1976, 
 \memras, 81, 89
\bibitem[Dotani et al.(1997)]{Dotani1997}
 Dotani, T., Yamashita, A., Ezuka, H., Takahashi, K., 
 Crew, G., Mukai, K., \& the SIS Team 1997, ASCA News, 5, 14
\bibitem[Filipovi\'{c} et al.(1998b)]{Filipovic1998b} 
 Filipovi\'{c}, M. D., Haynes, R. F., White, G. L., \& Jones, P. A.
 1998b, \aaps, 130, 421
\bibitem[Filipovi\'{c} et al.(1998a)]{Filipovic1998a} 
 Filipovi\'{c}, M. D., Pietsch, W., Haynes, R. F., White, G. L., 
 Jones, P. A., Wielebinski, R., Klein, U., Dennerl, K., Kahabka, P., 
 \& Lazendi\'{c}, J. S. 1998a, \aaps, 127, 119
\bibitem[Filipovi\'{c} et al.(2000a)]{Filipovic2000a}
 Filipovi\'{c}, M. D., Pietsch, W., \& Haberl, F. 
 2000a, \aap, 361, 823
\bibitem[Filipovi\'{c} et al.(2000b)]{Filipovic2000b}
 Filipovi\'{c}, M. D., Haberl, F., Pietsch, W., \& Morgan, D. H. 
 2000b, \aap, 353, 129
\bibitem[Finger, Wilson(2000)]{Finger2000}
 Finger, M. H., \& Wilson, C. A. 2000, \iaucirc, 7366
\bibitem[Gotthelf et al.(2000)]{Gotthelf2000}
 Gotthelf, E. V., Ueda, Y., Fujimoto, R., Kii, T., \& Yamaoka, K. 
 2000, \apj, 543, 417
\bibitem[Green(2000)]{Green2000}
`A Catalogue of Galactic Supernova Remnants (2000 August version)', 
Mullard Radio Astronomy Observatory, Cavendish Laboratory, Cambridge, 
United Kingdom (available on the World-Wide-Web at 
http://www.mrao.cam.ac.uk/surveys/snrs/)
\bibitem[Haberl et al.(1998)]{Haberl1998}
 Haberl, F., Angelini, L., Motch, C., \& White, N. E. 
 1998, \aap, 330, 189
\bibitem[Haberl et al.(2000)]{Haberl2000a}
 Haberl, F., Filipovic, M. D., Pietsch, W., \& Kahabka P. 
 2000, \aaps, 142, 41
\bibitem[Haberl, Sasaki(2000)]{Haberl2000b}
 Haberl, F., \& Sasaki, M. 2000, \aap, 359, 573
\bibitem[Hayashi et al.(1994)]{Hayashi1994}
 Hayashi, I., Koyama, K., Ozaki, M., Miyata, E., Tsunemi, H., 
 Hughes, J. P., \& Petre, R. 1994, \pasj, 46, L121
\bibitem[Henize(1956)]{Henize1956}
 Henize, K. G.\ 1956, \apjs, 2, 315
\bibitem[Hirayama et al.(1996)]{Hirayama1996}
 Hirayama, M., Nagase, F., Gunji, S., Sekimoto, Y., Saito, Y. 
 1996, ASCA News, 4, 18
\bibitem[Hughes(1994)]{Hughes1994b} 
 Hughes, J. P. 1994, \apj, 427, L25
\bibitem[Hughes, Smith(1994)]{Hughes1994a} 
 Hughes, J. P., \& Smith, R. C. 1994, \aj, 107, 1363
\bibitem[Imanishi et al.(1998)]{Imanishi1998} 
 Imanishi, K., Yokogawa, J., \& Koyama, K. 1998, \iaucirc, 7040
\bibitem[Imanishi et al.(1999)]{Imanishi1999}
 Imanishi, K., Yokogawa, J., Tsujimoto, M., \& Koyama, K. 
 1999, \pasj, 51, L15
\bibitem[Inoue et al.(1983)]{Inoue1983}
 Inoue, H., Koyama, K., \& Tanaka, Y. 1983, 
 in IAU Symposium 101,  
 Supernova Remnants and their X-Ray Emission, 
 ed. J. Danziger \& P. Gorenstein 
 (Dordrecht: D. Reidel Publishing Co.), 
 535 
\bibitem[In 't Zand et al.(2000)]{Intzand2000}
 In 't Zand, J. J. M., Halpern, J., Eracleous, M., 
 McCollough, M., Augusteign, T., Remillard, R. A., Heise, J. 
 2000, \aap, 361, 85
\bibitem[In 't Zand et al.(2001)]{Intzand2001}
 In 't Zand, J. J. M., Swank, J., Corbet, R. H. D., \& 
 Markwardt, C. B. 2001, \aap, 380, L26
\bibitem[Israel et al.(2000)]{Israel2000} 
 Israel, G. L., Campana, S., Covino, S., Dal Fiume, D., Gaetz, T. J., 
 Mereghetti, S., Oosterbroek, T., Orlandini, M., 
 et al. 2000, \apj, 531, L131
\bibitem[Israel et al.(1998b)]{Israel1998b} 
 Israel, G. L., Campana, S., Cusumano, G., Frontera, F., 
 Orlandini, M., Santangelo, A., \& Stella, L. 
 1998b, \aap, 334, L65
\bibitem[Israel et al.(1997)]{Israel1997} 
 Israel, G. L., Stella, L., Angelini, L., White, N. E., 
 Giommi, P., \& Covino, S. 
 1997, \apj, 484, L141
\bibitem[Israel et al.(1998a)]{Israel1998a} 
 Israel, G. L., Stella, L., Campana, S., Covino, S., Ricci, D., 
 \& Oosterbroek, T. 1998a, \iaucirc, 6999
\bibitem[Israel et al.(1999)]{Israel1999} 
 Israel, G. L., Stella, L., Covino, S., Campana, S., \& Mereghetti, S. 
 1999, \iaucirc, 7101
\bibitem[Kaaret et al.(1999)]{Kaaret1999}
 Kaaret, P., Piraino, S., Halpern, J., \& Eracleous, M. 1999, 
 \apj, 523, 197
\bibitem[Kahabka(2000)]{Kahabka2000}
 Kahabka, P. 2000, \aap, 354, 999
\bibitem[Kahabka, Pietsch(1996)]{Kahabka1996}
 Kahabka, P., \& Pietsch, W. 1996, \aap, 312, 919
\bibitem[Kahabka, Pietsch(1998)]{Kahabka1998}
 Kahabka, P., \& Pietsch, W. 1998, \iaucirc, 6840
\bibitem[Kohno et al.(2000)]{Kohno2000}
 Kohno, M., Yokogawa, J., \& Koyama, K. 
 2000, \pasj, 52, 299
\bibitem[Lamb et al.(2002b)]{Lamb2002}
 Lamb, R. C., Fox, D. W., Macomb, D. J., \& Prince, T. A.
 2002b, \apj, 574, L29
\bibitem[Lamb et al.(1999)]{Lamb1999}
 Lamb, R. C., Prince, T. A., Macomb, D. J. \& Finger, M. H. 
 1999, \iaucirc, 7081
\bibitem[Lamb et al.(2002a)]{Lamb2001}
 Lamb, R. C., Macomb, D. J., Prince, T. A., \& Majid, W. A. 
 2002a, \apj, 567, L129
\bibitem[Laycock et al.(2002)]{Laycock2002}
 Laycock, S., Corbet, R. H. D., Perrodin, D., Coe, M. J., 
 Marshall, F. E., \& Markwardt, C.
 2002, \aap, 385, 464
\bibitem[Liu et al.(2000)]{Liu2000}
 Liu, Q. Z., van Paradijs, J., \& van den Heuvel, E. P. J. 
 2000, \aaps, 147, 25
\bibitem[Liu et al.(2001)]{Liu2001}
 Liu, Q. Z., van Paradijs, J., \& van den Heuvel, E. P. J. 
 2001, \aap, 368, 1021
\bibitem[Lochner(1998)]{Lochner1998b} 
 Lochner, J. C. 1998, \iaucirc, 6858
\bibitem[Lochner et al.(1998)]{Lochner1998a} 
 Lochner, J. C., Marshall, F. E., Whitlock, L. A., \&  Brandt, N. 
 1998, \iaucirc, 6814
\bibitem[Lucke et al.(1976)]{Lucke1976}
 Lucke, R., Yentis, D., Friedman, H., Fritz, G., \& Shulman, S. 
 1976, \apj, 206, L25
\bibitem[Macomb et al.(1999)]{Macomb1999}
 Macomb, D. J., Finger, M. H., Harmon, B. A., Lamb, R. C., 
 \& Prince, T. A. 1999, \apj, 518, L99
\bibitem[Maragoudaki et al.(2001)]{Maragoudaki2001}
 Maragoudaki, F., Kontizas, M., Morgan, D. H., Kontizas, E., 
 Dapergolas, A., \& Livanou, E.\ 2001, \aap, 379, 864
\bibitem[Marshall et al.(1979)]{Marshall1979}
 Marshall, F. E., Boldt, E. A., Holt, S. S., Mushotzky, R. F., 
 Pravdo, S. H., Rothschild, R. E., \& Serlemitsos, P. J. 1979, \apjs, 40, 657
\bibitem[Marshall, Lochner(1998)]{Marshall1998}
 Marshall, F. E., \& Lochner, J. C. 1998, \iaucirc, 6818
\bibitem[Marshall et al.(1997)]{Marshall1997} 
 Marshall, F. E., Lochner, J. C., \& Takeshima, T. 1997, \iaucirc, 6777
\bibitem[Mathewson et al.(1983)]{Mathewson1983}
 Mathewson, D. S., Ford, V. L., Dopita, M. A., Tuohy, I. R., 
 Long, K. S., \& Helfand, D. J. 1983, \apjs, 51, 345
\bibitem[Mathewson et al.(1984)]{Mathewson1984}
 Mathewson, D. S., Ford, V. L., Dopita, M. A., Tuohy, I. R., 
 Mills, B. Y., \& Turtle, A. J. 1984, \apjs, 55, 189
\bibitem[Meaburn(1980)]{Meaburn1980}
 Meaburn, J. 1980, \mnras, 192, 365
\bibitem[Mereghetti et al.(2000)]{Mereghetti2000}
 Mereghetti, S., Tiengo, A., Israel, G. L., \& Stella, L. 2000, 
 \aap, 354, 567
\bibitem[Meyssonnier, Azzopardi(1993)]{Meyssonnier1993}
 Meyssonnier, N., \& Azzopardi, M. 1993, \aaps, 102, 451
\bibitem[Mitsuda et al.(1984)]{Mitsuda1984}
 Mitsuda, K., Inoue, H., Koyama, K., Makishima, K., Matsuoka, M., 
 Ogawara, Y., Shibazaki, N., Suzuki, K., Tanaka, Y., \& Hirano, T. 
 1984, \pasj, 36, 741
\bibitem[Motch et al.(1997)]{Motch1997}
 Motch, C., Haberl, F., Dennerl, K., Pakull, M., \& Janot-Pacheco, E. 
 1997, \aap, 323, 853
\bibitem[Murdin et al.(1979)]{Murdin1979}
 Murdin, P., Morton, D. C., \& Thomas, R. M.
 1979, \mnras, 1979, 186
\bibitem[Murphy, Bessell(2000)]{Murphy2000}
 Murphy, M. T., \& Bessell, M. S. 2000, \mnras, 311, 741
\bibitem[Nagase(1989)]{Nagase1989}
 Nagase, F. 1989, \pasj, 41, 1
\bibitem{Ohashi1996} 
 Ohashi, T., Ebisawa, K., Fukazawa, Y., Hiyoshi, K., Horii, M., 
 Ikebe, Y., Ikeda, H., Inoue, H., et al. 1996, \pasj\ 48, 157
\bibitem[Ozaki et al.(2000)]{Ozaki2000}
 Ozaki, M., Corbet, R. H. D., Marshall, F. E., \& Lochner, J. C. 
 2000, Adv.\ Space Res., 25, No.\ 3/4, 425
\bibitem[Paul et al.(2002)]{Paul2002}
 Paul, B., Nagase, F., Endo, T., Dotani, T., Yokogawa, J., \& 
 Nishiuchi, M. 2002, \apj, 579, 411
\bibitem[Petro et al.(1973)]{Petro1973}
 Petro, L., Feldman, F., \& Hiltner, W. A. 1973, \apj, 184, L123
\bibitem[Raymond, Smith(1977)]{Raymond1977}
 Raymond, J. C., \& Smith, B. W. 1977, \apjs, 35, 419
\bibitem[Reig et al.(1996)]{Reig1996}
 Reig, P., Chakrabarty, D., Coe, M. J., Fabregat, J., Negueruela, I.,
 Prince, T. A., Roche, P., \& Steele, I. A. 
 1996, \aap, 311, 879
\bibitem[Reig, Roche(1999)]{Reig1999}
 Reig, P., \& Roche, P. 1999, \mnras, 306, 100
\bibitem[Russell, Dopita(1992)]{Russell1992}
 Russell, S. C., \& Dopita, M. A. 1992, \apj, 384, 508
\bibitem[Sakano(2000)]{Sakano2000}
 Sakano, M. 2000, Ph.D. Thesis, Kyoto University
\bibitem[Sakano et al.(2000)]{Sakano2000b}
 Sakano, M., Torii, K., Koyama, K., Maeda, Y., \& Yamauchi, S. 
 2000, \pasj, 52, 1141
\bibitem[Santangelo et al.(1998)]{Santangelo1998} 
 Santangelo, A., Cusumano, G., Dal Fiume, D., Israel, G. L., 
 Stella, L., Orlandini, M., \& Parmar, A. N. 
 1998, \aap, 338, L59
\bibitem[Sasaki et al.(2001)]{Sasaki2001}
 Sasaki, M., Haberl, F., Keller, S., \& Pietsch, W. 
 2001, \aap, 369, L29
\bibitem[Sasaki et al.(2000)]{Sasaki2000}
 Sasaki, M., Haberl, F., \& Pietsch, W. 2000, \aaps, 147, 75
\bibitem[Schmidtke et al.(1999)]{Schmidtke1999} 
 Schmidtke, P. C., Cowley, A. P., Crane, J. D., Taylor, V. A., 
 McGrath, T. K., Huchings, J. D., \& Crampton, D. 
 1999, \aj, 117, 927
\bibitem[Serlemitsos et al.(1995)]{Serlemitsos1995}
 Serlemitsos, P. J., Jalota, L., Soong, Y., Kunieda, H., Tawara, Y., 
 Tsusaka, Y., Suzuki, H., Sakima, Y., et al. 1995, \pasj,  47, 105
\bibitem[Southwell, Charles(1996)]{Southwell1996}
 Southwell, K. A., \& Charles, P. A. 1996, \mnras, 281, L63
\bibitem[Stevens et al.(1999)]{Stevens1999}
 Stevens, J. B., Coe, M. J., \& Buckley, D. A. H. 
 1999, \mnras, 309, 421
\bibitem[Tanaka et al.(1994)]{Tanaka1994} 
 Tanaka, Y., Inoue, H., \& Holt, S. S. 1994, \pasj\ 46, L37
\bibitem[Telting et al.(1998)]{Telting1998}
 Telting, J. H., Waters, L. B. F. M., Roche, P., Boogert, A. C. A.,
 Clark, J. S., de Martino, D., \& Persi, P. 
 1998, \mnras, 296, 785
\bibitem[Torii et al.(2000b)]{Torii2000b} 
 Torii, K., Kohmura, T., Yokogawa, J., \& Koyama, K. 
 2000b, \iaucirc, 7441 
\bibitem[Torii et al.(2000a)]{Torii2000a} 
 Torii, K., Yokogawa, J., Imanishi, K., \& Koyama, K. 
 2000a, \iaucirc, 7428 
\bibitem[Tsujimoto et al.(1999)]{Tsujimoto1999} 
 Tsujimoto, M., Imanishi, K., Yokogawa, J., Koyama, K. 
 1999, \pasj, 51, L21 
\bibitem[Tucholke et al.(1996)]{Tucholke1996}
 Tucholke H.-J., de Boer K. S., \& Seitter W. C.
 1996, \aaps, 119, 91
\bibitem[Ueda et al.(1999)]{Ueda1999}
 Ueda, Y., Takahashi, T., Inoue, H., Tsuru, T., Sakano, M., 
 Ishisaki, Y., Ogasaka, Y., Makishima, K., et al. 1999, \apj, 518, 656
\bibitem[Ueno et al.(2000a)]{Ueno2000a} 
 Ueno, M., Yokogawa, J., Imanishi, K., \& Koyama, K. 
 2000a, \pasj, 52, L63
\bibitem[Ueno et al.(2000b)]{Ueno2000b} 
 Ueno, M., Yokogawa, J., Imanishi, K., \& Koyama, K.
 2000b, \iaucirc, 7442
\bibitem[van den Bergh(2000)]{Bergh2000}
 van den Bergh, S. 2000, \pasp, 112, 529
\bibitem[Vrtilek et al.(2001)]{Vrtilek2001}
 Vrtilek, S. D., Raymond, J. C., Boroson, B., 
 Kallman, T., Quaintrell, H., \& McCray, R. 2001, \apjl, 563, L139
\bibitem[Wang, Wu(1992)]{Wang1992}  
 Wang, Q., \& Wu, X. 1992, \apjs, 78, 391
\bibitem[e.g., Westerlund(1997)]{Westerlund1997}
 Westerlund, B. E. 1997, 
 in The Magellanic Clouds, 
 (New York: Cambridge University Press), 
 19 
%
\bibitem[White et al.(1983)]{White1983}
 White, N. E., Swank, J. H., \& Holt, S. S. 1983, \apj, 270, 711
\bibitem[Whitlock, Lochner(1994)]{Whitlock1994}
 Whitlock, L., \& Lochner, J. C. 1994, \apj, 437, 841
\bibitem[Williams et al.(1999)]{Williams1999}
 Williams, R.\ M., Chu, Y-H., Dickel, J.\ R.,
 Petre, R., Smith, R.\ C., \& Tavarez, M.\ 1999, \apjs, 123, 467
\bibitem[Wilson, Finger(1998)]{Wilson1998}
 Wilson, C. A., \& Finger, M. H. 1998, \iaucirc, 7048
\bibitem[Wojdowski et al.(1998)]{Wojdowski1998}
 Wojdowski, P., Clark, G. W., Levine, A. M., Woo, J. W., \& Zhang, S. N.
 1998, \apj, 502, 253
\bibitem[Ye et al.(1995)]{Ye1995} 
 Ye, T., Amy, S. W., Wang, Q. D., Ball, L., \& Dickel, J. 
 1995, \mnras, 275, 1218
\bibitem[Yokogawa(2002)]{Yokogawa2002b} 
 Yokogawa, J. 2002, Ph.D. thesis, Kyoto University, available at 
 $\langle$http://www-cr.scphys.kyoto-u.ac.jp/member/jun/job/phd/$\rangle$
\bibitem[Yokogawa et al.(2002)]{Yokogawa2001c} 
 Yokogawa, J., Imanishi, K., Koyama, K., Nishiuchi, M., \&
 Mizuno, N. 2002, \pasj, 54, 53
\bibitem[Yokogawa et al.(1999)]{Yokogawa1999} 
 Yokogawa, J., Imanishi, K., Tsujimoto, M., Kohno, M., \& Koyama, K. 
 1999, \pasj, 51, 547
\bibitem[Yokogawa et al.(2000e)]{Yokogawa2000e} 
 Yokogawa, J., Imanishi, K., Tsujimoto, M., Nishiuchi, M., 
 Koyama, K., Nagase, F., \& Corbet, R. H. D. 
 2000e, \apjs, 128, 491 
\bibitem[Yokogawa et al.(2000a)]{Yokogawa2000a} 
 Yokogawa, J., Imanishi, K., Ueno, M., \& Koyama, K. 
 2000a, \pasj, 52, L73 
\bibitem[Yokogawa et al.(2000d)]{Yokogawa2000d} 
 Yokogawa, J.,  Paul, B., Ozaki, M., Nagase, F., 
 Chakrabarty, D., \& Takeshima, T. 
 2000d, \apj, 539, 191 
\bibitem[Yokogawa et al.(2000b)]{Yokogawa2000b} 
 Yokogawa, J., Torii, K., Imanishi, K., \& Koyama, K. 
 2000b, \pasj, 52, L37 
\bibitem[Yokogawa et al.(2000c)]{Yokogawa2000c} 
 Yokogawa, J., Torii, K., Kohmura, T., Imanishi, K., \& Koyama, K. 
 2000c, \pasj, 52, L53
\bibitem[Yokogawa et al.(2001a)]{Yokogawa2001a} 
 Yokogawa, J., Torii, K., Kohmura, T., \& Koyama, K. 
 2001a, \pasj, 53, L9
\bibitem[Yokogawa et al.(2001b)]{Yokogawa2001b} 
 Yokogawa, J., Torii, K., Kohmura, T., \& Koyama, K. 
 2001b, \pasj, 53, 227
\bibitem[Yokogawa, Koyama(1998a)]{Yokogawa1998a} 
 Yokogawa, J., \& Koyama, K. 1998a, \iaucirc, 6853
\bibitem[Yokogawa, Koyama(1998b)]{Yokogawa1998b} 
 Yokogawa, J., \& Koyama, K. 1998b, \iaucirc, 7028
\bibitem[Yokogawa, Koyama(1998c)]{Yokogawa1998c} 
 Yokogawa, J., \& Koyama, K. 1998c, \iaucirc, 7009
\bibitem[Yokogawa, Koyama(2000)]{YK2000} 
 Yokogawa, J., \& Koyama, K. 2000, \iaucirc, 7361
\bibitem[Yoshii et al.(1996)]{Yoshii1996}
 Yoshii, Y., Tsujimoto, T., \& Nomoto, K. 
 1996, \apj, 462, 266
\end{thebibliography}
\end{document}